\documentclass[iop]{emulateapj}

\usepackage{graphicx}
\usepackage{psfig}
\usepackage{makeidx,amsmath,mathrsfs}
\usepackage{verbatim}
\usepackage{newtxtext,newtxmath}      
\usepackage[T1]{fontenc}
\usepackage{ae,aecompl}
\usepackage{amssymb}
\usepackage[dvipsnames]{xcolor}	
\usepackage{comment}
\usepackage{color}
\usepackage{lipsum}
\usepackage{pifont}
\usepackage{bm}
\usepackage{ulem}
\usepackage{balance}
\usepackage{multirow}
\usepackage{mathtools}
\usepackage{enumitem}
\usepackage{booktabs}
\usepackage{appendix}
\usepackage{listings}
\usepackage{hyperref}
\usepackage{lineno}
\hypersetup{
    colorlinks = true,
    citecolor = {MidnightBlue},
    linkcolor = {BrickRed},
    urlcolor = {BrickRed} 
}

\def\sec\ond{{\rm s}}

\def\be{\begin{equation}}\def\bea{\begin{eqnarray}}\def\beaa{\begin{eqnarray*}}
\def\ee{\end{equation}}  \def\eea{\end{eqnarray}}  \def\eeaa{\end{eqnarray*}}

\shorttitle{Critical Points and Massive Neutrinos}
\shortauthors{J. Moon, G. Rossi, H. Yu (2022)}
\begin{document}
\title{Signature of Massive Neutrinos from the Clustering of Critical Points: \\ I. Density-threshold-based Analysis in Configuration Space}
\author{Jeongin Moon$^1$, Graziano Rossi$^{1,2,*}$, Hogyun Yu$^1$}
\affil{$^1$ Department of Physics and Astronomy, Sejong University, Seoul, 143-747, Republic of Korea  \\
         $^2$ Max Planck Institute for Astrophysics, Karl-Schwarzschild-Str. 1, D-85748 Garching, Germany}
\email{$^{*}$Corresponding Author: Graziano Rossi (graziano@sejong.ac.kr)}



\begin{abstract}


Critical points represent a subset of special points tracing cosmological structures, carrying remarkable topological properties. 
They thus offer a richer high-level description of the \textit{multiscale} cosmic web, being more robust to systematic effects. For the first time, 
we characterize here their clustering statistics in massive neutrino cosmologies, including cross-correlations, 
and quantify their simultaneous imprints on the corresponding web constituents -- i.e., halos, filaments, walls, and voids -- for a series of rarity levels. 
Our first analysis is centered on a density-threshold-based  approach in configuration space. 
In particular,  we show that the presence of massive neutrinos does affect the 
baryon acoustic oscillation peak amplitudes of all of 
the critical point correlation functions above/below the rarity threshold, 
as well as the positions of  their correspondent inflection points at large scales:
departures from analogous measurements carried out 
in the baseline massless neutrino scenario can reach up to
$\sim 7\%$ in autocorrelations and $\sim 9\%$ in cross-correlations
at $z=0$ when $M_{\nu}=0.1~{\rm eV}$, and are more pronounced for higher neutrino mass values. 
In turn, these combined \textit{multiscale} effects can be used as a novel technique to set upper limits on
the summed neutrino mass and infer the type of hierarchy. 
Our study is particularly 
relevant for ongoing and future large-volume 
redshift surveys such as the Dark Energy Spectroscopic Instrument and the Rubin Observatory Legacy Survey of Space and Time, 
which will provide unique datasets suitable for establishing competitive neutrino mass constraints.

\end{abstract}






\section{Introduction} \label{sec_introduction}
 

Neutrinos represent a clear indication of physics beyond the standard model, 
since the confirmation via oscillation experiments that they are massive particles -- 
see, e.g., \cite{LesgourguesPastor2006} and 
\cite{Gonzalez-Garcia2008} for seminal reviews on the phenomenology of massive neutrinos.
In this respect, the baseline $\Lambda$CDM cosmological framework characterized by 
massless neutrinos (or at best by a minimal non-zero neutrino mass), a
spatially flat cosmology dominated by collisionless cold dark matter (CDM), and a dark 
energy (DE) component in the form of a cosmological constant ($\Lambda$),  
should be extended accordingly. 

Therefore, it comes with no surprise that 
determining the neutrino mass scale and type of
hierarchy are among the major enterprises 
of all of the ongoing and upcoming 
large-volume astronomical surveys, 
and one of the primary targets of future space missions. 
This is for example the case of
the Dark Energy Survey  \citep[DES;][]{DES2005}, 
the Dark Energy Spectroscopic Instrument \citep[DESI;][]{DESICollaboration2016a}, 
the Rubin Observatory Legacy Survey of Space and Time  \citep[LSST;][]{LSST2019}, 
the Prime Focus Spectrograph \citep[PFS;][]{Takada2014},
the Nancy Grace Roman Space Telescope \citep[Roman;][]{Spergel2015},
and the Euclid Consortium \citep{Laureijs2011}, 
as well as of next-generation cosmic microwave background (CMB) experiments 
such as CMB Stage 4 \citep[CMB-S4;][]{Abazajian2016, Abazajian2019, Abitbol2017}
or the Simons Observatory \citep{Simons2019}, in combination with 21-cm surveys 
like the Square Kilometre Array (SKA).\footnote{\url{https://skatelescope.org}} 

To this end, cosmology is mainly sensitive to the 
summed neutrino mass -- denoted as $M_{\nu} \equiv \sum_{\rm i} m_{\rm i}$ throughout this paper,
with $m_{\rm i}$ ($i=1,2,3$) representing the three individual  masses of active neutrinos -- and provides competitive upper bounds on $M_{\nu}$.
On the other hand, particle physics experiments set stringent lower bounds
(i.e.,  $M_{\nu} \ge 0.05{~\rm eV}$). Moreover, 
flavor oscillations bring no information on the absolute
neutrino mass scale and are insensitive to individual neutrino masses,
as they allow only for a measurement of 
squared differences. Namely, 
$\Delta m^2_{21} = m_2^2 - m_1^2 \simeq + 7.50 \times 10^{-5} {\rm eV}^2$ can be obtained from solar neutrino oscillations, and
$\Delta m^2_{31} = m_3^2 - m_1^2 \simeq \Delta m^2_{32} = m_3^2 - m_2^2 
\simeq \pm 2.45 \times 10^{-3} {\rm eV}^2$ can be inferred from atmospheric neutrino oscillations \citep{Gonzalez-Garcia2008}. 
 Since we only know experimentally 
that $\Delta m_{21} > 0$, as well as the value of
$|\Delta m_{31}|$, this 
leads to the possibility of either
normal hierarchy (NH; $m_1 < m_2 \ll m_3$) 
or inverted hierarchy (IH; $m_3 \ll m_1  \simeq m_2$) for the mass ordering. 
In the NH configuration, the minimal sum of neutrino masses 
is $M_{\nu} = 0.057 ~{\rm eV}$, while in the IH configuration 
the minimal summed mass is $M_{\nu} = 0.097~{\rm eV}$. 

The latest neutrino mass upper bounds reported by the 
Planck collaboration 
are $M_{\nu} < 0.12~{\rm eV}$ \citep{Planck2020cosmo}, 
while the most constraining combination of data 
including Data Release 16 (DR16) from 
the extended Baryon Oscillation Spectroscopic Survey  
\citep[eBOSS;][]{Dawson2016}, part of the fourth generation of the 
Sloan Digital Sky Survey \citep[SDSS-IV;][]{York2000, Blanton2017}, 
gives the upper limit on the sum of neutrino masses at  $M_{\nu} < 0.115~{\rm eV}$ \citep{eBOSS2021}. 
Clearly, such stringent bounds put considerable pressure on the IH scenario
as a plausible possibility for the neutrino mass ordering. 


A direct neutrino mass detection, or at least more  competitive upper limits on $M_{\nu}$
are expected in the next few years from Stage-IV cosmological experiments. 
For example, 
DESI  will be able to measure $M_{\nu}$
with an uncertainty of $0.020~{\rm eV}$
for $k_{\rm max} < 0.2 h {\rm Mpc}^{-1}$,  
sufficient to make the first direct detection 
of $M_{\nu}$ at more than $3\sigma$ significance
and rule out the IH at the 99\%
confidence level (CL) if the hierarchy is normal and the masses are minimal \citep{DESICollaboration2016a}.
Moreover,  future CMB-S4 measurements combined with late-time measurements of galaxy clustering and cosmic shear from the 
Rubin Observatory LSST would allow one to
achieve a $3\sigma$ 
detection of the minimal mass sum $M_{\nu}^{\rm min} =0.06~{\rm eV}$
\citep{MishraSharma2018}. 
In synergy with neutrino $\beta$ decay experiments 
sensitive to
the electron neutrino mass
\citep[i.e., KATRIN;][]{KATRIN2001}, and neutrinoless double $\beta$ 
decay probes sensitive to the effective Majorana mass 
\citep[i.e., KamLAND, GERDA, Cuore;][]{Gando2013,Ackermann2013,CUORE2015}, 
it should be possible to finally close on  
the neutrino mass scale and type of
hierarchy within this decade. 
For extensive details to this end, see, i.e., \cite{Gariazzo2018}.


Constraints on massive neutrinos from cosmology can be obtained 
exploiting a variety of tracers and probes, spanning different scales.
The traditional and perhaps 
most straightforward route is via the CMB, 
typically involving CMB gravitational lensing  or using the 
early integrated Sachs-Wolfe (ISW) effect in polarization maps
-- see, e.g.,  \cite{Battye2014},  \cite{Brinckmann2019}, \cite{Planck2020lensing}, and references therein. 
Regarding baryonic tracers of the large-scale structure (LSS),
a plethora of techniques have been proposed in the literature to date.  
Among them, a (largely incomplete) recent list 
includes three-dimensional (3D) galaxy clustering via the 
matter power spectrum \citep{Angulo2021,Bose2021},
bispectrum \citep{Hahn2020,Hahn2021}, 
and marked power spectrum  \citep{Massara2021},
cosmic shear, peaks, and bispectrum through weak lensing \citep{Coulton2019,Ajani2020}, 
galaxy clusters via the Sunyaev-Zel'dovich (SZ) effect \citep{Roncarelli2017,Bolliet2020},
peculiar velocities \citep{Whitford2022},
voids statistics \citep{Kreisch2019,Zhang2020}, 
the Lyman-$\alpha$ (Ly$\alpha$) forest flux power spectrum \citep{Seljak2005,Seljak2006,Viel2010,Rossi2017,Rossi2020},
and much more. 
The  common denominator among all these different methodologies
is to rely on a single observable, and/or on a specific scale (or a limited scale-range). 
Only recently, there have been some interesting attempts 
to use instead combinations of probes -- see for example the work of \cite{Bayer2021}.


Furthermore,  
gaining a deeper theoretical understanding
of the impact of massive neutrinos on structure formation (particularly at small scales)
and on the primary observables typically used to characterize  neutrino mass effects
are necessary, to obtain robust
constraints free from systematic biases. 
Consistent progress has been
made in this direction over the last few years.
Relevant examples include
the detailed assessment of 
neutrino effects on Ly$\alpha$ forest observables \citep{Rossi2017},
the quantification of the consequences of non-zero neutrino masses and asymmetries on dark matter halo assembly bias 
\citep{Lazeyras2021, Wong2022}, the implications of 
massive neutrinos in the realm of the Hubble tension \citep{DiValentino2022},
the challenges related to parameter degeneracies when combining CMB and LSS probes in massive
neutrino cosmologies 
\citep{Archidiacono2017}, and
the  estimation of the impact of massive neutrinos on the  baryon acoustic oscillation (BAO)
peak \citep{Peloso2015} and 
on the linear point (LP) of the spatial correlation function \citep{ Parimbelli2021}.


On an apparently unrelated subject, 
recently an interesting analysis of the cosmic
web in terms of \textit{critical points} 
(i.e.,  a subset of special points in position-smoothing space, tracing cosmological structures)
has been proposed by
\cite{Cadiou2020}\footnote{Note though that their primary focus is on critical events, not to be confused with critical points.}, 
and the corresponding clustering properties of those points 
have been characterized in \cite{Shim2021}
and in \cite{Kraljic2022} 
as a function of rarity threshold
in the baseline $\Lambda$CDM model.
Critical points (extrema and saddles, where the spatial gradient of the density field vanishes) carry remarkable topological 
properties, and provide a more fundamental
view of the  cosmic web as a whole -- in a \textit{multiscale} perspective. 
This is primarily because the 
topology of the initial density field, at a 
fixed smoothing scale, is encoded in the positions and heights of such points: 
hence, in principle, it it possible to 
predict the evolution of 
cosmic web structures 
at later times from their clustering properties jointly with the  power spectrum of the underlying initial Gaussian field. 
Moreover, the special scales at which two critical points coalesce produce merging effects
corresponding to the mergers of halos, filaments, walls, and voids \citep{Cadiou2020}.
In addition, the drift of critical points with smoothing defines the skeleton tree 
\citep{Sousbie2008,Pogosyan2009,Gay2012}, capturing topological 
variations with increasing smoothing scale -- that acts effectively as a time variable. 
Besides being more robust to systematic effects, 
critical points are thus useful because 
they represent a meaningful and efficient compression
of information of the 3D density field, capturing its 
most salient features -- as they carry
significance in terms of cosmology and/or
galaxy formation.  In fact, with some
care (see our discussion later on in Section \ref{sec_results}), critical points can be associated
to corresponding physical structures of the same type, although their typical size will depend
on the smoothing scale of the density field.  
In particular, and of direct relevance to this work, 
\cite{Shim2021} found an amplification of the BAO features with increasing  density threshold 
in the auto-correlations of critical points (reversed for cross-correlations) within the $\Lambda$CDM, and reported of
an `\textit{inflection scale}' around $\sim 133h^{-1}{\rm Mpc}$ which appears to be 
common to all of the  auto- and cross-correlation functions. 

 
Here, we combine all of the  previous (seemingly unrelated) topics in a
coherent framework, and characterize, for the first time, 
the clustering statistics of critical points in massive neutrino
cosmologies --  addressing their sensitivity  to small 
neutrino masses, 
with the goal of identifying 
multiscale signatures.    
Specifically, we compute the auto- and cross-correlation functions of critical points
in configuration space, for a 
series of rarity thresholds, and also quantify redshift evolution effects. 
Our main analysis is centered on BAO scales, and it is focused on two
key aspects: (1) the multiscale effects of massive neutrinos on the 
BAO
peak amplitudes of all of 
the critical point correlation functions above/below rarity threshold; and, (2)
the multiscale effects of massive neutrinos on the spatial positions of 
 their correspondent correlation function inflection points at large scales.\footnote{See
Section \ref{subsec_auto2pcf_cp}
for the definition of such inflection points at large spatial separations.} 
The first aspect is inferred from the fact that 
massive neutrinos do impact the
BAO peak \citep{Peloso2015}, while the second  one
is inspired by a sort of resemblance (although in a \textit{multiscale} perspective)
with  the  LP of the spatial correlation function, 
which is also affected by a non-zero neutrino mass \citep{ Parimbelli2021}. 


We carry out our measurements using 
a subset of realizations from the \texttt{QUIJOTE} suite \citep{VillaescusaNavarro2020}, as
explained in Section \ref{sec_simulations}. In particular,
we utilize full snapshots at redshifts $z=0,1,2,3$ for 
a choice of representative neutrino masses, in order to assess redshift evolution effects. 
Our first analysis is centered on a `\textit{density threshold-based}' approach: 
the methodology to construct density fields from
the output of $N$-body simulations, extract and classify critical points, and
perform clustering measurements in configuration space is thoroughly 
explained in Section  \ref{sec_methodology}. 
Noticeably, in this study we consider three cuts in rarity $\mathcal{R}$, namely 
$\mathcal{R} \in \{5, 10, 20 \}$ -- where  the
abundances are expressed in percentages.
 
 
Our main results are detailed in Section \ref{sec_results}.
Here, we show that the presence of massive neutrinos 
do affect the BAO peak amplitudes of all of 
the critical point correlation functions above/below rarity threshold, 
as well as the positions of  their correspondent  inflection points at large scales.
Departures from analogous measurements carried out 
in the baseline massless neutrino scenario can reach up to
$\sim7\%$ in auto-correlations
and  $\sim9\%$ in cross-correlations
at $z=0$ when $M_{\nu}=0.1~{\rm eV}$, and are more 
pronounced for higher neutrino mass values. 
In a companion publication,  we show how
these combined \textit{multiscale} effects derived from the clustering of critical points 
can be used as a novel powerful technique (complementary to more traditional methods) to set 
upper limits on $M_{\nu}$ and infer the type of hierarchy. 


This work is the first of a series of investigations that
aim at exploring the sensitivity of critical points and critical events 
to massive neutrinos, and more generally in relation to the dark sector. 
The layout of the paper is organized as follows. 
Section \ref{sec_simulations}  briefly describes the
simulations considered in our study.
Section \ref{sec_methodology} presents the
methodology at the core of
the `\textit{density threshold-based}' approach.
The main results of our analysis are provided in 
Section   \ref{sec_results}, including cross-correlation measurements as well as
the characterization of redshift evolution effects.  
We conclude in Section \ref{sec_conclusions}, where we
summarize the various results and highlight novel aspects -- along with ongoing applications and future avenues. 
We leave in Appendix \ref{sec_appendix_A} a number of technicalities 
on the analysis performed, related in particular to our choices of 
bin size and  smoothing scale of the density field,  and provide 
some complementary tables in Appendix \ref{sec_appendix_B}.



\section{Simulations} \label{sec_simulations}

 


\begin{table}
\centering
\caption{Characteristics of the simulations used in this study.}
\doublerulesep2.0pt
\renewcommand\arraystretch{1.5}
\begin{tabular}{cc} 
\hline \hline   
 \hspace{1.5cm} \texttt{QUIJOTE} Suite \\
\hline
{ \hspace{1.5cm} \it Relevant Parameters} \\
\hline
$\Omega_{\rm m}$ &         0.3175\\
$\Omega_{\rm \Lambda}$ &  0.6825    \\
$\Omega_{\rm b}$ &            0.0490   \\
$n_{\rm s}$  &                 0.9624\\
$h$             &          0.6711\\ 
$w$             &        -1 \\
$M_{\nu}$ [eV]              &     0.0, 0.1, 0.4\\
$A_{\rm S}$ [baseline]     & 2.13 $\times$ $10^{-9}$  \\  
$\sigma_8 (z=0) $   &  0.834 \\
$k_0$   [Mpc$^{-1}$] & 0.05 \\  
\hline
{ \hspace{1.5cm} \it Simulation Details} \\
\hline
Box [$h^{-1}$Mpc] & 1000 \\
$N_{\rm CDM}$ &   $512^3$  \\
$N_{\nu}$ &   $512^3$    \\
$z_{\rm in}$ &  127  \\
ICs Type &  Zel'dovich \\
$z$ used & 0,1,2,3 \\
\hline
\hline
\end{tabular}
\label{table_sims_details}
\end{table}
 

\begin{figure*}
\centering
\includegraphics[angle=0,width=0.95\textwidth]{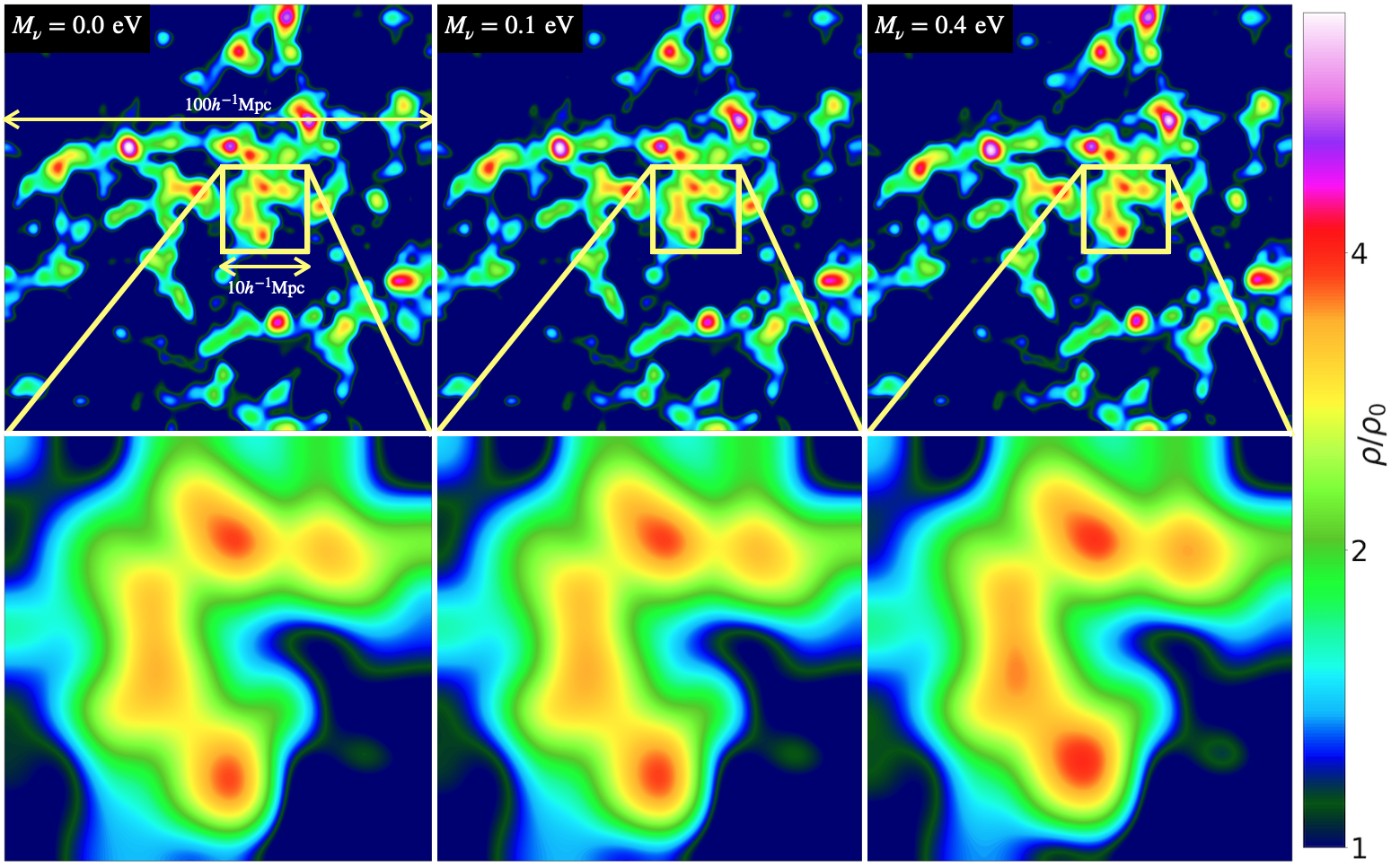} 
\caption{Spatial projection of the DM overdensity field from three selected \texttt{QUIJOTE} simulations  at $z=0$.
Top panels display $100 \times 100~[h^{-1} {\rm Mpc}]^2 $ slices of 
the simulation cubes; 
bottom panels are enlargements of a 
small $10 \times 10~[h^{-1} {\rm Mpc}]^2$ inset with identical spatial location. 
The fiducial massless neutrino cosmology (left), a
massive neutrino model with $M_{\nu}=0.1~{\rm eV}$ (middle),
and a massive neutrino scenario with $M_{\nu}=0.4~{\rm eV}$ (right) are shown, respectively. 
Although tiny, morphological and topological differences induced by a non-zero neutrino mass on the cosmic web network 
-- as traced by the large-scale density field  -- are visually perceptible.}
\label{fig_visualization_sims}
\end{figure*}


In this work, we use a 
subset of realizations from the \texttt{QUIJOTE} suite.
The \texttt{QUIJOTE} simulations \citep{VillaescusaNavarro2020} are a set of 44,100 
publicly available $N$-body runs\footnote{\url{https://quijote-simulations.readthedocs.io/en/latest/}}
spanning over 7000 cosmological models, following the gravitational evolution 
of $N_{\rm CDM}$ CDM plus $N_{\nu}$
neutrino particles (if present) with 
different resolutions (i.e., $256^3$, $512^3$, $1024^3$ particles per species). 
Initial conditions (ICs) are set at redshift $z_{\rm in}=127$
either adopting the Zel'dovich 
approximation \citep{Zeldovich1970} or via second order Lagrangian Perturbation Theory (2LPT).
Moreover, the \texttt{QUIJOTE} suite contains \textit{standard},
\textit{fixed}, and \textit{paired fixed} simulations, the difference being in the way
ICs are generated. 
The input matter power spectra and transfer functions are obtained
via the Boltzmann code \texttt{CAMB} \citep{Lewis2000}.
All of the realizations are produced  with the
 Three Particle Mesh (TreePM) code
\texttt{GADGET-III} \citep{Springel2005}, over a 
periodic box size of 1$h^{-1}$Gpc, for a 
total volume of $44,100$ ($h^{-1}$Gpc)$^3$.
Snapshots are saved at $z=0, 0.5, 1, 2, 3$, respectively. 
The gravitational softening is set to $1/40$ of the mean inter-particle 
distance for all of the particle species, and 
the ICs random seeds are the same
for an identical realization in different models,
but vary from realization to realization within the same model.
The  baseline cosmological parameters
are closer to those reported by the  Planck Collaboration in 2018 \citep{Planck2020cosmo},
representing a flat (i.e., $\Omega_{\Lambda} + \Omega_{\rm m} = 1$) massless neutrino $\Lambda$CDM model
with 
the total matter density $\Omega_{\rm m} =0.3175$, the DE density parameter $\Omega_{\Lambda} = 0.6825$, 
the baryon density $\Omega_{\rm b} = 0.0490$,
the primordial scalar spectrum power-law index  $n_{\rm s} = 0.9624$, 
the Hubble parameter  $h=0.6711$, 
the DE equation of state (EoS) parameter $w=-1$,
and the primordial power spectrum amplitude $A_{\rm s} = 2.13 \times 10^{-9}$ 
at the pivot scale
$k_{\rm pivot} \equiv k_0 =0.05~{\rm Mpc^{-1}}$.  
For convenience, these reference parameters are reported in 
the upper part of 
Table \ref{table_sims_details}. 
Note that the normalization choice for $A_{\rm s}$  in the
baseline massless neutrino cosmology implies 
$\sigma_8=0.834$ at $z=0$, with 
$\sigma_8$ the 
linear theory root mean square (\textit{rms}) matter fluctuation in $8h^{-1} {\rm Mpc}$ spheres;
by construction, all of the other 
realizations are  
tuned to match an identical $\sigma_8$ value at the present time.  
 

In  massive neutrino cosmologies, the rescaling method 
developed by \cite{Zennaro2017} is used to determine initial conditions.
In addition, all of the massive neutrino realizations assume degenerate masses:
the summed neutrino mass is indicated  
with $M_{\nu}$ in Table \ref{table_sims_details}. 
Regarding neutrino implementations within the $N$-body framework, 
all of the the neutrino runs are  
performed employing a popular particle-based approach, treated as a 
 collisionless and pressureless fluid (in the same fashion as CDM).
Moreover, the forces at small scales for the neutrino species are also properly computed. 
The particle-based approach, despite shot noise challenges, is
automatically able to capture the full nonlinear neutrino clustering 
and it accurately reproduces the nonlinear evolution at 
small scales -- a crucial aspect for obtaining reliable and robust cosmological constraints. 
See \cite{Rossi2017,Rossi2020} for additional details on this implementation. 


In our study, we use 100 
independent realizations (i.e., having different initial random seeds) from the \texttt{QUIJOTE} suite 
for any of the following cosmologies: 
fiducial framework (massless neutrino scenario or `best guess'), 
massive neutrino model with $M_{\nu}=0.1~{\rm eV}$, and
massive neutrino model with $M_{\nu}=0.4~{\rm eV}$,
respectively. 
Specifically, we exploit the runs denoted as \textit{standard},
characterized by Zel'dovich ICs at $z=127$.
The resolutions considered are relatively low, $512^3$ particles per species, 
on a periodic box size of 1$h^{-1}$Gpc. 
We utilize full snapshots at $z=0,1,2,3$, in order to also assess redshift evolution effects. 
As previously noted, $\sigma_8(z=0)=0.834$ in all of the runs;
hence, $A_{\rm s}$ varies with different neutrino masses.  
This normalization convention is what has been termed
`NORM' in \cite{Rossi2020}, and is 
well-motivated  observationally, since $\sigma_8$ at the present epoch
is effectively dictated by observational constraints.  
The main characteristics of the simulations adopted here
are reported in the bottom part of Table \ref{table_sims_details},
and more technical details can also be found in \cite{VillaescusaNavarro2020}. 


As a visual example,  Figure \ref{fig_visualization_sims} shows 
a spatial projection of the dark matter (DM) overdensity field 
 along the $x-y$ plane 
with a depth of $50 h^{-1}{\rm Mpc}$ across the $z$-axis, 
obtained from three selected \texttt{QUIJOTE} simulations   at $z=0$.
From left to right, the plots refer to the 
fiducial massless neutrino cosmology run ($M_{\nu}=0.0~{\rm eV}$), the 
massive neutrino model with $M_{\nu}=0.1~{\rm eV}$, and the 
massive neutrino scenario with $M_{\nu}=0.4~{\rm eV}$, respectively. 
Top panels display  $100 \times 100~[h^{-1} {\rm Mpc}]^2 $
slices of the simulation cubes,
while bottom panels are enlargements of a 
small $10 \times 10 ~[h^{-1} {\rm Mpc}]^2$ inset with identical spatial location 
and the same thickness as in the top figures.  
The normalized density fields are rendered using a rainbow palette,
with the corresponding values indicated by the right side color bar. 
While rather small, 
morphological and topological differences induced by a non-zero neutrino mass
are visually perceptible: they are quantifiable via
the clustering statistics of critical points in the complex structure of the evolving cosmic web network, 
as we show in Section \ref{sec_results}.   
 


\section{Methodology}  \label{sec_methodology}


In this section, we briefly describe our
technique to construct density fields from
the output of $N$-body simulations, extract and classify critical points, and
subsequently perform clustering measurements in configuration space.  We generically refer to this methodology as
the `\textit{density threshold-based}' approach. 


\subsection{Density Threshold-Based Approach: Overview}  

 
For this first work, we  adopt a 
technique centered on the extraction and classification of critical points 
from smoothed density fields. The construction of such
fields is obtained via a grid interpolation
strategy, directly from the DM
particle distribution of $N$-body simulations. 
In essence, in the first step DM particles are turned into cells of 
given density, that are properly smoothed up to a desired scale. 
Critical points are then identified, and classified according to their type. Subsequently, those points are  
sorted in terms of density threshold (or rarity), 
and for a fixed rarity their clustering statistics is characterized as a function
of $M_{\nu}$ 
and redshift. 
The advantage of this three-step procedure resides in its
simplicity, ease of implementation, good efficiency, and 
effective numerical performance.
In what follows, we
briefly describe the three main building blocks of our pipeline. 


\subsection{Construction of Density Fields}  


We begin with the production of density fields, which are 
constructed from the DM particle output of 
selected \texttt{QUIJOTE} $N$-body realizations using
the Python libraries for the analysis of numerical simulations
\citep[\texttt{Pylians};][]{VillaescusaNavarro2018}.\footnote{See \url{https://github.com/franciscovillaescusa/Pylians}}
\texttt{Pylians} are a set of \texttt{Python}, \texttt{Cython}, and \texttt{C} libraries 
that are helpful in the analysis of large-volume runs, providing a number of 
statistical tools -- as well as routines
to create and manipulate density fields. 
Specifically, at a fixed redshift, we build 3D density fields  $\rho(\mathbf{x})$  
from the spatial positions ($\mathbf{x}$) of $N$-body particles
using a triangular-shape cloud
(TSC) mass-assignment scheme
to deposit particle masses into the grid;
in this process, no weights are associated to each particle.
Density  contrast fields,
defined as 
$\delta(\mathbf{x}) = \rho(\mathbf{x}) / \bar{\rho} -1$ 
with $\bar{\rho}$ the corresponding average background density, 
are readily inferred from the global knowledge of $\rho(\mathbf{x})$.
We adopt $N_{\rm grid} = 1024$ voxels per dimension on a regular grid spacing -- corresponding to $2N_{\rm p}$, with 
$N_{\rm p}=512$ the number of particles per dimension in a given \texttt{QUIJOTE} simulation snapshot. 
We also tested different mass-assignment schemes, as well as a number of voxel choices for $N_{\rm grid}$,
and found negligible dependence on our results. 
 

\subsection{Extraction and Classification of Critical Points}  \label{sec_methodology_cp_classification}


In the next step,
fields are subsequently smoothed 
before the extraction and classification of critical points.  More explicitly, 
DM density fields created from \texttt{QUIJOTE} snapshots are
smoothed with a Gaussian kernel $W_{\rm G}$ over $R_{\rm G}=3h^{-1}{\rm Mpc}$ scales,
using \texttt{Py-Extrema}.\footnote{See \url{https://github.com/cphyc/py_extrema}}
Smoothing  
is relevant for this
\textit{density threshold-based} methodology: we provide more extensive details related to our choice of the smoothing scale in 
Appendix \ref{sec_appendix_A}.
We will also return on this important point in a companion publication,
where we confront the present technique with a (fundamentally different) persistence-based approach -- which instead does not rely directly on smoothing.
In essence, the effect of smoothing is 
analogous to a resolution cut-off, erasing small-scale structures: 
at a given $z$-interval, structures of different sizes and masses are
present, and therefore selecting an explicit $R_{\rm G}$ would correspond
to averaging over specific 
density regions.
A suitable smoothing choice is then relevant
if we were to identify critical points with actual physical structures,
as the effective typical size of such objects depends on the
adopted smoothing scale of the density field: 
to this end, see the considerations presented in Section \ref{subsec_visualizations_cp}.

Once density fields have been properly smoothed, critical points are
identified with the same detection algorithm adopted in \cite{Cadiou2020} and in \cite{Shim2021},
based on a local quadratic estimator centered on 
a second order Taylor expansion of the density field about critical points -- see also \cite{Gay2012} for additional details.
Such expansion yields:
\begin{equation} 
\mathbf{x} - \mathbf{x_{\rm c}} \sim (\nabla \nabla \rho)^{-1} \nabla \rho,
\label{eq_cp}
\end{equation}
with $\mathbf{x_{\rm c}}$ the spatial position of the 
critical point.
In essence, for each grid cell the 
gradient and Hessian of the density field are 
computed, Equation (\ref{eq_cp})
is solved, and solutions found at a distance greater than 1 pixel 
are discarded via
the requirement ${\rm max}_{\rm i}(|\mathbf{x}_{\rm i} - \mathbf{x}_{\rm c} |) < 1~{\rm pixel}$ --
with $i=1,2,3$:
only the critical point closest to the center of the cell 
is retained.
Subsequently, each cell enclosing
a critical point is flagged, 
and an efficient loop is performed 
over flagged cells containing multiple critical points of the same type.
In this process, critical points are classified according to their corresponding rank order.
Specifically, the classification is 
based on the number of negative eigenvalues of the density 
Hessian
-- i.e., $0$ for voids (${\mathcal V}$ or minima);  $1$ for walls (${\mathcal W}$ or `saddle 1' or wall-type saddles); $2$ for filaments (${\mathcal F}$ 
or `saddle 2' or filament-type saddles);
$3$ for peaks (${\mathcal P}$ or maxima).\footnote{Throughout the paper, we will 
refer to a specific critical  point via its type, although the reader should 
keep in mind the important remark detailed in Section \ref{subsec_visualizations_cp}:
the correspondence between density-based critical points and cosmological structures depends on the smoothing scale choice.}


\subsection{Clustering Measurements in Configuration Space}  \label{sec_methodology_clustering}


At this stage of the pipeline, critical points have been properly identified and classified.
Next,  all of the points are sorted according to the chosen density threshold $\nu = \delta / \sigma$,
with $\delta$ the overdensity previously defined (but now referred to the \textit{smoothed} density field), and $\sigma$
the corresponding \textit{rms} fluctuation of the field. 
Following  \cite{Shim2021}, we adopt an identical 
\textit{ad hoc} convention
in defining rarity levels:  namely, we always sample the population that
provides the same abundance for a given type of critical point.
This choice represents also a relevant point, that should be
kept in mind later on, as it carries some impact on the interpretation of the results presented in
Section \ref{sec_results} 
(i.e., the adopted rarity definition allows one to 
avoid density overlapping among critical points, thus enhancing more remarkable features in correlations).

Our rarity level definition can be formalized as follows. Let's first 
introduce the notation $\alpha \in \{0,1,2,3\} \equiv \{{\mathcal V}, {\mathcal W}, {\mathcal F}, {\mathcal P}\}$ 
to specify the critical point type. We will also use the indices $\ell$ and $k$ in place of $\alpha$, whenever necessary (i.e., $\alpha \equiv \ell, k$). 
In essence, 
for peak- and filament-type (respectively, void- and wall-type) critical points,
we select ensambles with $\nu$ higher (respectively, lower) than a given threshold $\nu_{\alpha, \mathcal{R}}$,
where  $\mathcal{R}$ indicates the relative abundance cut.\footnote{Note that throughout the paper
we will refer  indistinctly to either $\nu_{\alpha, \mathcal{R}}$ or $\mathcal{R}$ to indicate the rarity level/threshold, 
with the latter quantity defined as in Equations 
\ref{eq_abundance_up} and \ref{eq_abundance_down} and expressed in percentage terms.}
The threshold is then fixed as  the rarity for each type of critical point $\alpha$
yielding the same relative abundances defined by the ratios:
\begin{equation} 
 {n_{\rm C, \alpha}(\nu \ge \nu_{\alpha, \mathcal{R}})  \over n_{\rm C, \alpha}}  
\label{eq_abundance_up}
\end{equation}  
with $\alpha \in \{\mathcal{P},\mathcal{F}\} \equiv \{3,2\}$  
for peaks and filaments, respectively,
and: 
\begin{equation} 
 {n_{\rm C, \alpha}(\nu \le \nu_{\alpha, \mathcal{R}})  \over n_{\rm C, \alpha}}
\label{eq_abundance_down}
\end{equation}    
with $\alpha \in \{\mathcal{W},\mathcal{V}\} \equiv \{1,0\}$  
for walls and voids, respectively. 
In the previous expressions,
$n_{\rm C, \alpha}$ indicates the entire ensemble of critical points of type $\alpha$,
while $n_{\rm C, \alpha}(\nu \ge \nu_{\alpha, \mathcal{R}})$
or $n_{\rm C, \alpha}(\nu \le \nu_{\alpha, \mathcal{R}})$
represent the subset containing 
critical points of type $\alpha$ above or below the
selected threshold $\nu_{\alpha, \mathcal{R}}$.
For ease of notation, in what follows we 
define $\gamma \equiv \nu_{\alpha, \mathcal{R}}$,
and also indicate $n_{\rm C, \alpha}(\nu \ge \gamma)$
or $n_{\rm C, \alpha}(\nu \le \gamma)$  
generically with $n_{\rm C, \alpha}^{\gamma}$. 
In our study we consider three cuts in rarity, namely 
$\mathcal{R} \in \{5, 10, 20 \}$ -- where  the
abundances are expressed in percentages.

We then compute the spatial clustering statistics of  critical points above (below) threshold -- i.e.,  real space two-point auto- and cross-correlations $\xi(r)$
as a function of the separation $r$ --
using the \textsc{Halotools}\footnote{See \url{https://github.com/astropy/halotools}}  
platform \citep{Hearin2017},
equipped with
efficient algorithms for calculating clustering statistics (including cross-correlations). 
For our two-point calculations in configuration space, we
adopt the widely used minimal-variance Landy-Szalay (LS) estimator \citep{LandySzalay1993}
assuming periodic boundary conditions, 
and utilize random catalogs always at least 20 times bigger in size than the considered data.\footnote{We have also 
extensively tested the impact of increasing the size of random catalogs
in this process, and found negligible effects on our results.} 
Specifically, for a given catalog $C^{\gamma}_{\alpha}$ of critical points above (below) a 
selected density threshold $\gamma$ 
having total size $n^{\gamma}_{\rm C, \alpha}$, and assuming a corresponding random catalog 
$R^{\gamma}_{\alpha}$ of total size $n^{\gamma}_{\rm R, \alpha}$ characterized by a random uniform probability
distribution of points within the same volume, 
the LS estimator for the two-point correlation function 
$\xi_{\rm k \ell}^{\rm \gamma, \rm LS}$ above (below)
$\gamma$ reads\footnote{Throughout the rest of the paper, for ease of simplicity, we omit to indicate 
the threshold levels 
$\gamma \equiv \nu_{\alpha, \mathcal{R}}$ in the notation of the two-point correlation functions. The adopted rarity choices will be readily distinguishable
in the various plots presented,  via the usage of different line styles, point shapes, or contrasting colors.}:
\begin{equation}
\xi_{\rm k \ell}^{\rm \gamma, \rm LS} =  \sqrt{\eta \beta} {\langle C^{\gamma}_{\rm k}  C^{\gamma}_{\rm \ell} \rangle \over \langle R^{\gamma}_{\rm k}  R^{\gamma}_{\rm \ell} \rangle}  -
\sqrt{\eta} {\sqrt{    \langle C^{\gamma}_{\rm k}  R^{\gamma}_{\rm \ell} \rangle     \langle C^{\gamma}_{\rm \ell}  R^{\gamma}_{\rm k} \rangle       } \over   \langle R^{\gamma}_{\rm k}  R^{\gamma}_{\rm \ell} \rangle }   +1
\label{eq_cf_ls}
\end{equation}
where the normalization factors are specified by 
\begin{equation}
\eta =  {(n^{\gamma}_{\rm R, \rm k} -1) (n^{\gamma}_{\rm R, \rm \ell} -1) \over  n^{\gamma}_{\rm C, \rm k} n^{\gamma}_{\rm C, \rm \ell}}
\end{equation}
and 
\begin{equation}
\beta =  {n^{\gamma}_{\rm R, \rm k} n^{\gamma}_{\rm R, \rm \ell} \over (n^{\gamma}_{\rm C, \rm k} -1) (n^{\gamma}_{\rm C, \rm \ell} -1)}.
\end{equation}
Note that all of the correlation   measurements obtained via Equation (\ref{eq_cf_ls}) are reported as a function of the
spatial separation  $r$, conventionally expressed in units of $h^{-1}{\rm Mpc}$. 
Unless specified otherwise, we always
use a bin size of $5h^{-1}{\rm Mpc}$: this choice is motivated by the study presented 
in Appendix \ref{sec_appendix_A}, where we also discuss smoothing and bin size effects on clustering measurements.
Moreover, in order to accurately determine 
the spatial locations of the inflection points of the correlation functions at large scales
(see Section \ref{subsec_auto2pcf_cp}), 
we also  employ a `refinement' technique where we adopt a smaller bin size of
$1h^{-1}{\rm Mpc}$ in selected regions near such points.



\section{Density Threshold-Based Approach: Results} \label{sec_results}


This section contains the main results of our first analysis.
After some basic considerations related to the 
`\textit{density threshold-based}' approach,
we present here auto- and cross-correlation measurements
of the clustering of critical points in massive neutrino cosmologies for a series of rarity levels, confronted
with analogous computations in the baseline massless neutrino model.
In particular, we show  
how massive neutrinos  
affect the BAO peak amplitudes of all of 
the critical point correlation functions above/below rarity threshold, 
as well as the positions of  their corresponding 
inflection points at large scales. 
Finally, we also address redshift evolution effects.

In what follows,  unless specified otherwise, we always express rarity thresholds $\mathcal{R}$
 in percentage and consider two massive neutrino models having $M_{\nu}=0.1~{\rm eV}$
and $M_{\nu}=0.4~{\rm eV}$, respectively, besides
the  baseline massless neutrino scenario.
Moreover, all of the measurements  represent
averages over 100 independent \texttt{QUIJOTE} realizations at fixed cosmology,
and the associated errorbars are the
corresponding $1\sigma$ variations.


\subsection{Abundance and Visualization of Critical Points in Massive Neutrino Cosmologies}  \label{subsec_visualizations_cp}


\begin{figure*}
\centering
\includegraphics[angle=0,width=0.97\textwidth]{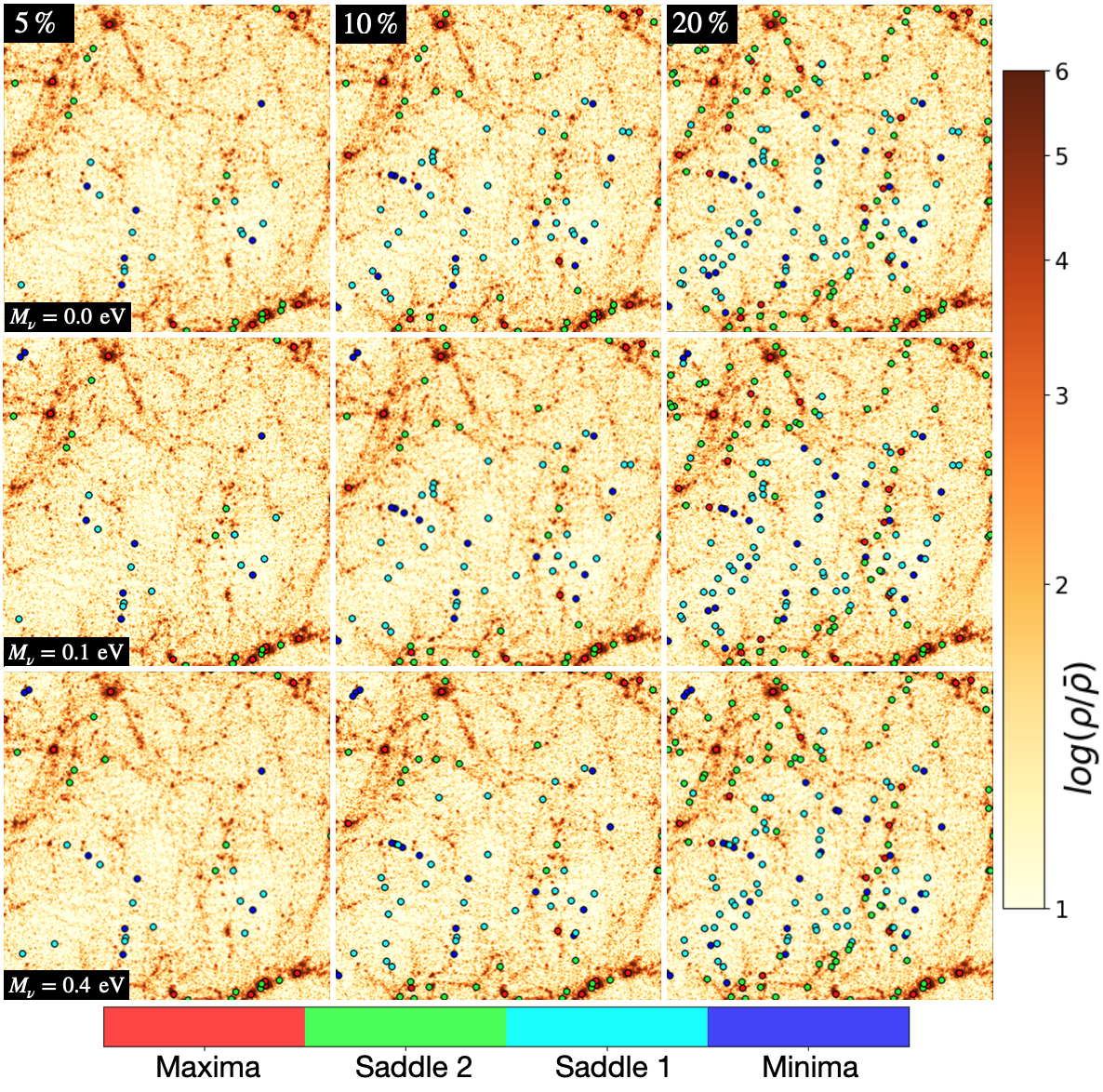} 
\caption{Visualization of the spatial distribution of critical points (color-coded by their corresponding type) in massless and massive neutrino cosmologies at $z=0$, from small
 \texttt{QUIJOTE}-simulated cubic density patches of $100 h^{-1}{\rm Mpc}$ side and $ 50 h^{-1}{\rm Mpc}$ depth, as a function of rarity
 -- with $\mathcal{R}$ increasing from left to right, as indicated in the various panels. 
Red, green, cyan, and blue are used respectively
to display maxima, filament-saddles (`saddle 2'),
wall-saddles (`saddle 1'), and minima.
 A `\textit{density threshold-based}' approach is adopted for the identification and classification of critical points, as reported in Section \ref{sec_methodology}. See the main text for more details.}
\label{fig_visualization_critical_points}
\end{figure*}


We begin with an instructive 
visualization of the geometrical distribution of critical points identified and classified
via the `\textit{density threshold-based}' procedure described in Section \ref{sec_methodology}.
Figure \ref{fig_visualization_critical_points} shows the 
spatial location of critical points  
in \texttt{QUIJOTE}-simulated patches at $z=0$, color-coded by type, for three different rarity thresholds. 
The squared patches in the figure are characterized by a side of 
 $100 h^{-1}{\rm Mpc}$ in the $x-y$ plane, and a $50 h^{-1}{\rm Mpc}$ depth along the $z$-coordinate. 
The underlying normalized density fields are 
the results of sampling the 
density fields with \texttt{Pylians}
assuming $N_{\rm grid}=1024$.
Top panels refer to the baseline cosmology (massless neutrino scenario),
middle panels are for a massive neutrino cosmology with $M_{\nu}=0.1~{\rm eV}$,
while bottom panels display results for $M_{\nu}=0.4~{\rm eV}$. 
From left to right, the rarity threshold is increased in terms of
relative abundance cut, corresponding to $5\%$, $10\%$, and $20\%$, respectively,
via the selection criteria detailed in Section \ref{sec_methodology_clustering}. 

As anticipated before (i.e., Section \ref{sec_methodology_cp_classification}),
care must be taken in directly identifying 
critical points derived from a 
`\textit{density threshold-based}' approach
with cosmological structures.   
This is primarily because such methodology
relies on smoothing, and  
therefore 
the effective typical size of the LSS cosmic web constituents 
 depends on the
 smoothing scale of the density field, which acts as a 
resolution cut-off.
To this end, see also the discussion in \cite{Shim2021}, who adopted a smoothing scale
of $R_{\rm G} = 6h^{-1}{\rm Mpc}$ corresponding to
$\sim 10^{15} {\rm M_{\odot}}$ average density regions.
As noted by the same authors, 
with such a choice of $R_{\rm G}$
only large voids can be resolved, and 
only 5\% of all density peaks
represent virialized galaxy clusters at $z=0$, while the rest is still in a collapse process.
From Figure \ref{fig_visualization_critical_points}, 
one can deduce that decreasing the rarity threshold (namely, moving from the left to the right panels in the figure) 
provides a more detailed mapping of the corresponding
underlying cosmological structures.
This fact 
can be inferred from correlation function measurements:
larger values of $\mathcal{R}$ 
manifest into smaller dispersions/scatter in $\xi(r)$,
as we show in Sections \ref{subsec_auto2pcf_cp} and \ref{subsec_crosscorr_cp}.  
Moreover, it is also interesting to compare (even just at the visual level),
the spatial distribution of critical 
points for a fixed rarity value with increasing neutrino mass: 
differences in the abundance and spatial location of critical points when $M_{\nu} \ne 0.0~{\rm eV}$
translate into 
morphological and topological differences that 
carry rich cosmological information.  

In this regards, Table \ref{table_abundance_cp} reports the
overall abundance of 
the entire set of critical points at $z=0$,
classified by type, in the three cosmological frameworks considered in this work. 
These measurements are the results of 
running \texttt{Py-Extrema} on the 
smoothed density fields, and 
 represent  averages over 100 independent \texttt{QUIJOTE} realizations at fixed cosmology. 
Note that for a Gaussian field it is 
expected that the number of peaks ($\mathcal{P}$) and voids ($\mathcal{V}$)  is similar, as well as the
number of filaments ($\mathcal{F}$)  versus walls ($\mathcal{W}$).
Moreover, in Gaussian fields, the ratio between filaments and peaks ($\mathcal{F/P}$)
or walls and voids ($\mathcal{W/V}$)
is estimated to be $\sim 3.05$, and in addition 
the total number of extrema and saddle points 
is preserved at the first non-Gaussian perturbative order. 
Also,  for sufficiently large volumes, the ratio between the number of peaks
and walls over voids and filaments ($[\mathcal{P/W}]/[\mathcal{V/F}]$) in a Gaussian field 
is very close to unity throughout the entire $z$-evolution. 
This is because the genus topology, equal to the
alternate sum of critical points, should be preserved, namely:
\begin{equation}
n_{\rm cp}^{\mathcal{P}}  - n_{\rm cp}^{\mathcal{F}}  + n_{\rm cp}^{\mathcal{W}}  - n_{\rm cp}^{\mathcal{V}}  =0~,
\end{equation} 
with $n_{\rm cp}^{\rm (i)}$ the mean number density of peaks, filaments, walls, and voids -- respectively. 

However, nonlinear evolution breaks the symmetry between underdense and overdense regions.
Table \ref{table_abundance_cp_Gauss}
summarizes all of the previous considerations, showing clear departures from Gaussianity --
expected because of nonlinear structure formation, in addition to the presence of massive neutrinos. 
Note that the effect of a non-zero neutrino mass 
is represented by an overall decrement in the abundance of critical points, much more pronounced for
higher neutrino masses. For peaks, this fact has 
repercussions at the level of the halo mass function,
and it can be interpreted in the framework of the halo model -- see e.g. \cite{Rossi2017}, their Section 4.3. 
 
 
\begin{table}
\centering
\caption{Total number of critical points at $z=0$, classified by type, for the three cosmologies considered in this work.}
\doublerulesep2.0pt
 \begin{tabular}{c|c|c}
\hline \hline   
   & ${\mathbf{M_{\nu}{\rm [eV]}}}$ & ${\mathbf{z=0}}$  \\ 
\hline \hline  
&  0.0 & $228,094 \pm 382$  \\
\cline{2-2}
{\bf Minima ($\mathcal{V}$)} & 0.1  & $227,770 \pm 355$  \\
\cline{2-2}
&  0.4 & $226,527 \pm 402$ \\
\hline
&  0.0 & $680,088 \pm 823$  \\
\cline{2-2}
{\bf Saddle 1 ($\mathcal{W}$)} & 0.1 & $678,915 \pm 775$  \\
\cline{2-2}
&  0.4 & $674,157 \pm 797$  \\
\hline
&  0.0 & $669,208 \pm 858$  \\
\cline{2-2}
{\bf Saddle 2 ($\mathcal{F}$)} & 0.1  & $667,952 \pm 826$   \\
\cline{2-2}
&  0.4 & $662,960 \pm 766$  \\
\hline
&  0.0 & $223,951 \pm 337$  \\ 
\cline{2-2}
{\bf Maxima ($\mathcal{P}$)} & 0.1 & $223,571 \pm 319$  \\
\cline{2-2}
&  0.4 & $222,031 \pm 325$  \\
\hline \hline
\end{tabular}
\label{table_abundance_cp} 
\end{table}

 
\begin{table}
\centering
\caption{Relevant abundance ratios of critical points at $z=0$ for the three cosmologies adopted in this study, highlighting the fact that nonlinear evolution as well as
the effects of massive neutrinos (if present) break the
symmetry between overdense and underdense regions -- expected instead in a purely Gaussian field.}
\doublerulesep2.0pt
  \begin{tabular}{c|c|c}
\hline \hline   
   & ${\mathbf{M_{\nu}{\rm [eV]}}}$ & ${\mathbf{z=0}}$  \\ 
\hline \hline  
&  0.0 & $0.9818 \pm 0.0022$ \\
\cline{2-2}
{\bf  $\mathcal{P/V}$} & 0.1 & $0.9816 \pm 0.0021$ \\
\cline{2-2}
&  0.4 & $0.9802 \pm 0.0023$  \\
\hline
&  0.0 & $0.9840 \pm 0.0017$ \\
\cline{2-2}
{\bf $\mathcal{F/W}$} & 0.1 & $0.9839 \pm 0.0017$    \\
\cline{2-2}
&  0.4 & $0.9834 \pm 0.0016$   \\
\hline
&  0.0 & $2.9888 \pm 0.0059$  \\
\cline{2-2}
{\bf $\mathcal{F/P}$} & 0.1 & $2.9877 \pm 0.0056$  \\
\cline{2-2}
&  0.4 & $2.9859 \pm 0.0056$   \\
\hline
&  0.0 & $2.9816 \pm 0.0062$   \\ 
\cline{2-2}
{\bf $\mathcal{W/V}$} & 0.1 & $2.9807 \pm 0.0058$   \\
\cline{2-2}
&  0.4 & $2.9761 \pm 0.0063$  \\
\hline
&  0.0 & $0.9661 \pm 0.0028$   \\ 
\cline{2-2}
{\bf ($\mathcal{P/W}$)/($\mathcal{V/F}$)} & 0.1 & $0.9657 \pm 0.0026$  \\
\cline{2-2}
&  0.4 & $0.9639 \pm 0.0027$  \\
\hline \hline
\end{tabular}
\label{table_abundance_cp_Gauss} 
\end{table}

 

\subsection{Auto-Correlations of Critical Points in Massive Neutrino Cosmologies}  \label{subsec_auto2pcf_cp}
 
 
\begin{figure*}
\centering
\includegraphics[angle=0,width=0.90\textwidth]{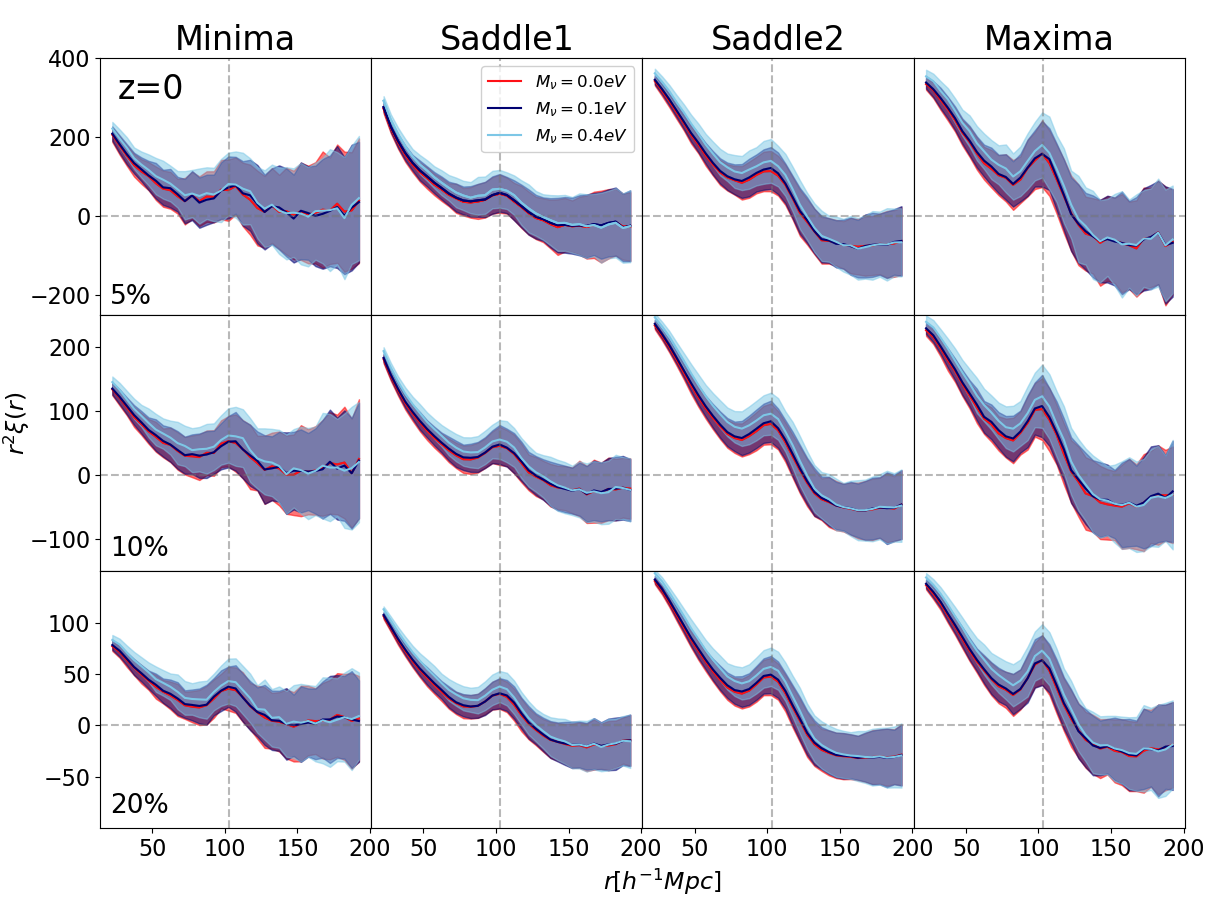} 
\caption{Auto-correlations of critical points in configuration space
at $z=0$, for three different rarity thresholds: $\mathcal{R}=5$ (top panels),
$\mathcal{R}=10$ (middle panels), $\mathcal{R}=20$ (bottom panels).
The two-point clustering statistics of minima, 
wall-type saddles, 
filament-type saddles, 
and maxima 
are shown from left to right, respectively. 
Three different cosmological models are considered, as indicated in the 
figure with contrasting 
colors: $M_{\nu}=0.0~{\rm eV}$ (red),
$M_{\nu}=0.1~{\rm eV}$ (blue), and $M_{\nu}=0.4~{\rm eV}$ (cyan).
 Errorbars are estimated from averages of 
100 independent \texttt{QUIJOTE} realizations for a given cosmology.  
BAO peaks are clearly visible in all of the panels, 
with the vertical dashed grey lines marking their exact scale  ($102.5h^{-1}{\rm Mpc}$) -- which is independent 
of critical point type, neutrino mass, and rarity.}
\label{fig_auto_clustering}
\end{figure*} 


Next, we move to auto-correlation measurements.
Figure \ref{fig_auto_clustering} shows the
clustering statistics of critical points in configuration space
at $z=0$ for three different rarity thresholds ($\mathcal{R} =5,10,20$), 
computed with the LS estimator via Equation (\ref{eq_cf_ls}).
From left to right, minima, 
wall-type saddles, 
filament-type saddles, 
and  maxima 
are displayed, 
respectively,
 for the three cosmologies considered in our analysis.
BAO peaks are clearly visible in all of the panels, 
with the vertical dashed grey lines marking their exact scale (expressed with $r$ in $h^{-1}{\rm Mpc}$). 
Note that the positions of the BAO peaks are all the same, independently of critical point type, neutrino mass, and rarity -- namely $102.5h^{-1}{\rm Mpc}$,
which precisely coincides with the BAO expected location in the reference cosmology.   
This remarkable aspect is a clear indication that critical points trace the BAO peak similarly to DM, halos, and galaxies, and are
faithful representations of their complementary structures.

Several interesting features can be inferred from Figure \ref{fig_auto_clustering}. 
First, it is evident that  
a non-zero neutrino mass generally corresponds to higher BAO amplitudes, regardless of the specific critical point type, 
and those amplitudes increase as $M_{\nu}$ is augmented. Namely,
the larger the neutrino mass, the 
bigger the BAO amplitudes; this 
implies that it may be possible to infer neutrino mass signatures from such differential measurements, 
as we argue in this section. 
Moreover, the corresponding BAO amplitudes of minima and wall-type saddles are 
smaller than those of maxima and filament-type saddles.
The similarity in shape between minima and saddles-1
(respectively, maxima and saddles-2) 
can be explained by considering how critical points below (above) threshold are
selected -- as detailed in Section \ref{sec_methodology_clustering}.
In fact, for minima and wall-type saddles the specific critical point abundance 
is determined from the  
lower-limit of the density field, while the opposite
is done for  maxima and filament-type saddles.
Generally, extrema (i.e., minima and maxima) 
show noisier and less smoother correlation function shapes when compared
to saddle-point clustering: notice in fact
that their corresponding errorbars in Figure \ref{fig_auto_clustering} are much larger. 
This is readily explained by the fact that 
the time evolution of extrema is more nonlinear than that of saddle points,  
as also reported by \cite{Cadiou2020} and \cite{Kraljic2022}.
In addition, more marked features in the BAO peaks and noisier correlation function shapes
are seen with smaller rarity thresholds (see for example the top panels of the figure, where $\mathcal{R}=5$).
This finding can be simply interpreted via linear bias, meaning that the more the tracer gets biased, the
stronger the clustering  -- e.g., \cite{Kaiser1984,Desjacques2018,Shim2021,Kraljic2022}.
Note that the effect of a rarity cut is similar to that of smoothing
(see Appendix \ref{sec_appendix_A}, and also the previous references):
an increase in smoothing implies a decrease in the number of volume 
elements along with an increase in bias, and consequently the spatial correlation function 
gets noisier because of enlarged statistical uncertainties, and 
its features appear more enhanced.   

\begin{figure}
\centering
\includegraphics[angle=0,width=0.495\textwidth]{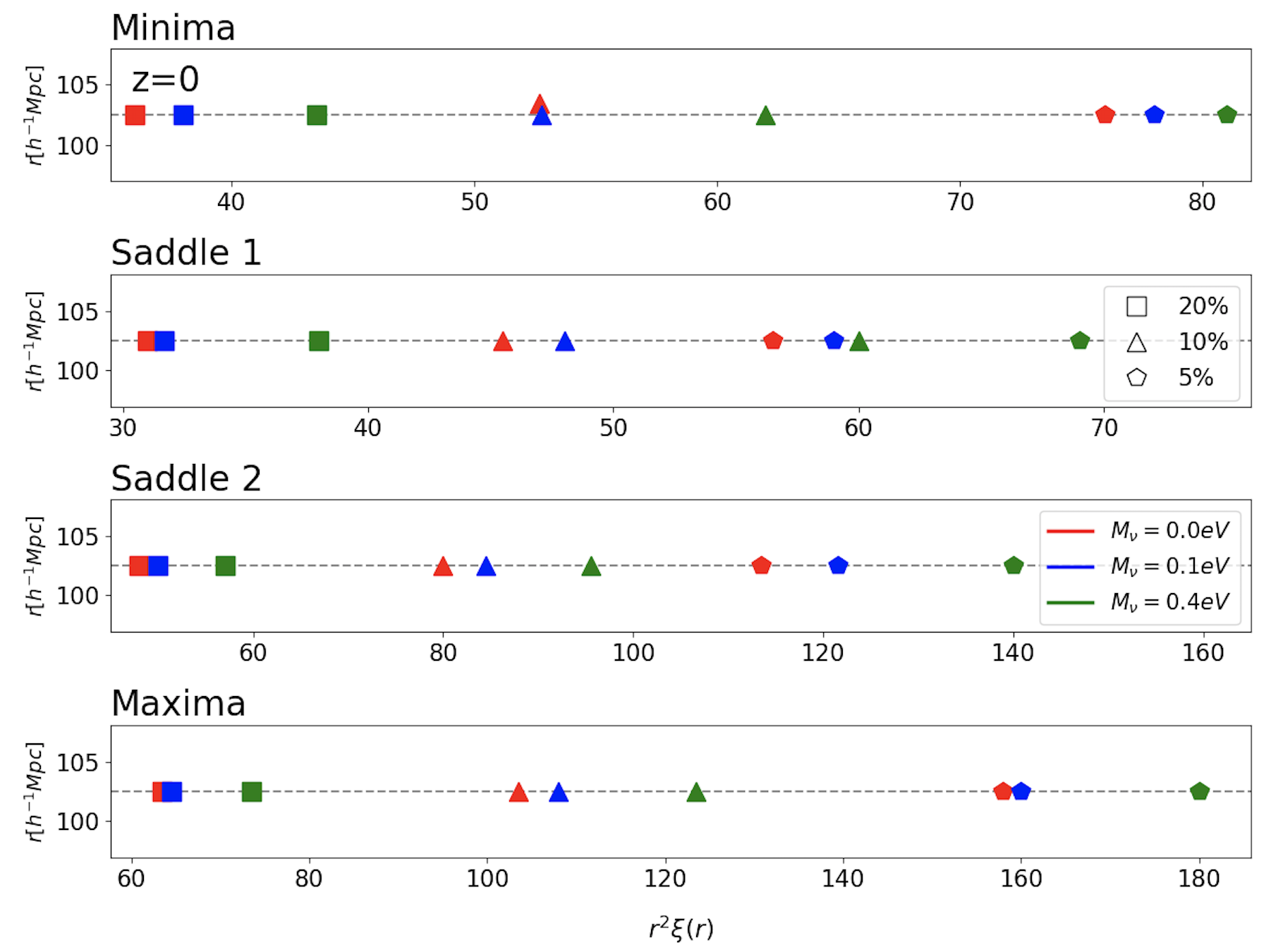} 
\caption{BAO amplitudes ($r^2\xi$) versus their correspondent  spatial position (i.e., $102.5h^{-1}{\rm Mpc}$, independent of critical point type, neutrino mass, and rarity) at $z=0$,
derived from the auto-correlation measurements of minima, saddles-1, saddles-2, and maxima (top to bottom panels) reported in Figure \ref{fig_auto_clustering}. 
The three cosmological scenarios considered are represented with different colors (red: $M_{\nu}=0.0~{\rm eV}$; blue: $M_{\nu}=0.1~{\rm eV}$;  green: $M_{\nu}=0.4~{\rm eV}$),
and the rarity thresholds are indicated via contrasting symbols ($\mathcal{R}=5$, pentagons; $\mathcal{R}=10$, triangles; $\mathcal{R}=20$, squares).}
\label{fig_auto_bao_amplitude}
\end{figure}

\begin{figure}
\centering
\includegraphics[angle=0,width=0.495\textwidth]{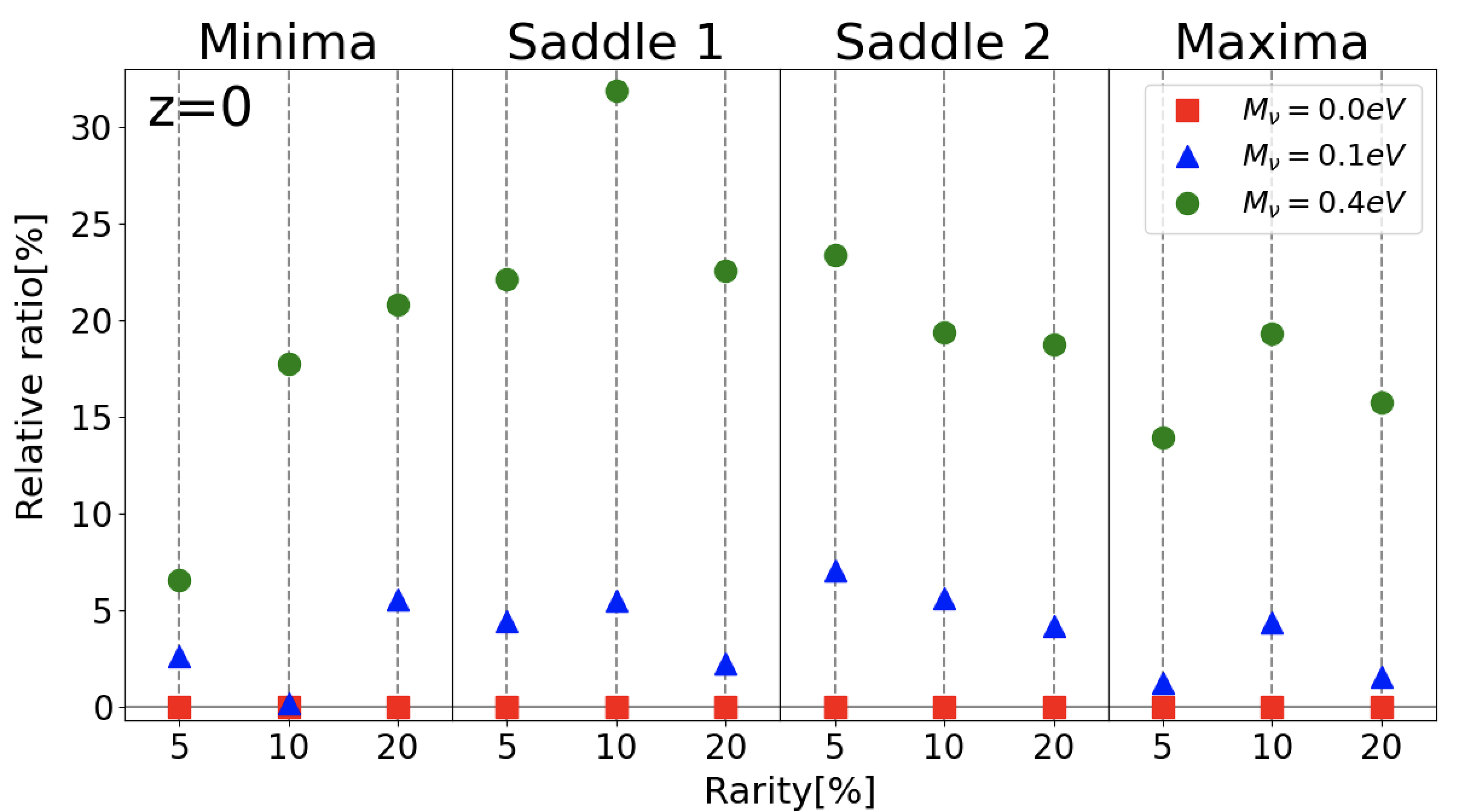}
\caption{BAO amplitudes in massive neutrino cosmologies, normalized by their correspondent values in a massless neutrino scenario and expressed in percentages 
as a function of rarity threshold at $z=0$. Colors refer to the same cosmological models examined in Figure  \ref{fig_auto_bao_amplitude}. 
Departures from analogous measurements carried out in the baseline $M_{\nu}=0.0~{\rm eV}$ framework can reach up to 
$\sim 7\%$ at $z = 0$, even for a relatively small neutrino mass of $M_{\nu}=0.1~{\rm eV}$.}
\label{fig_auto_bao_amplitude_ratios}
\end{figure}

\begin{figure*}
\centering
\includegraphics[angle=0,width=0.85\textwidth]{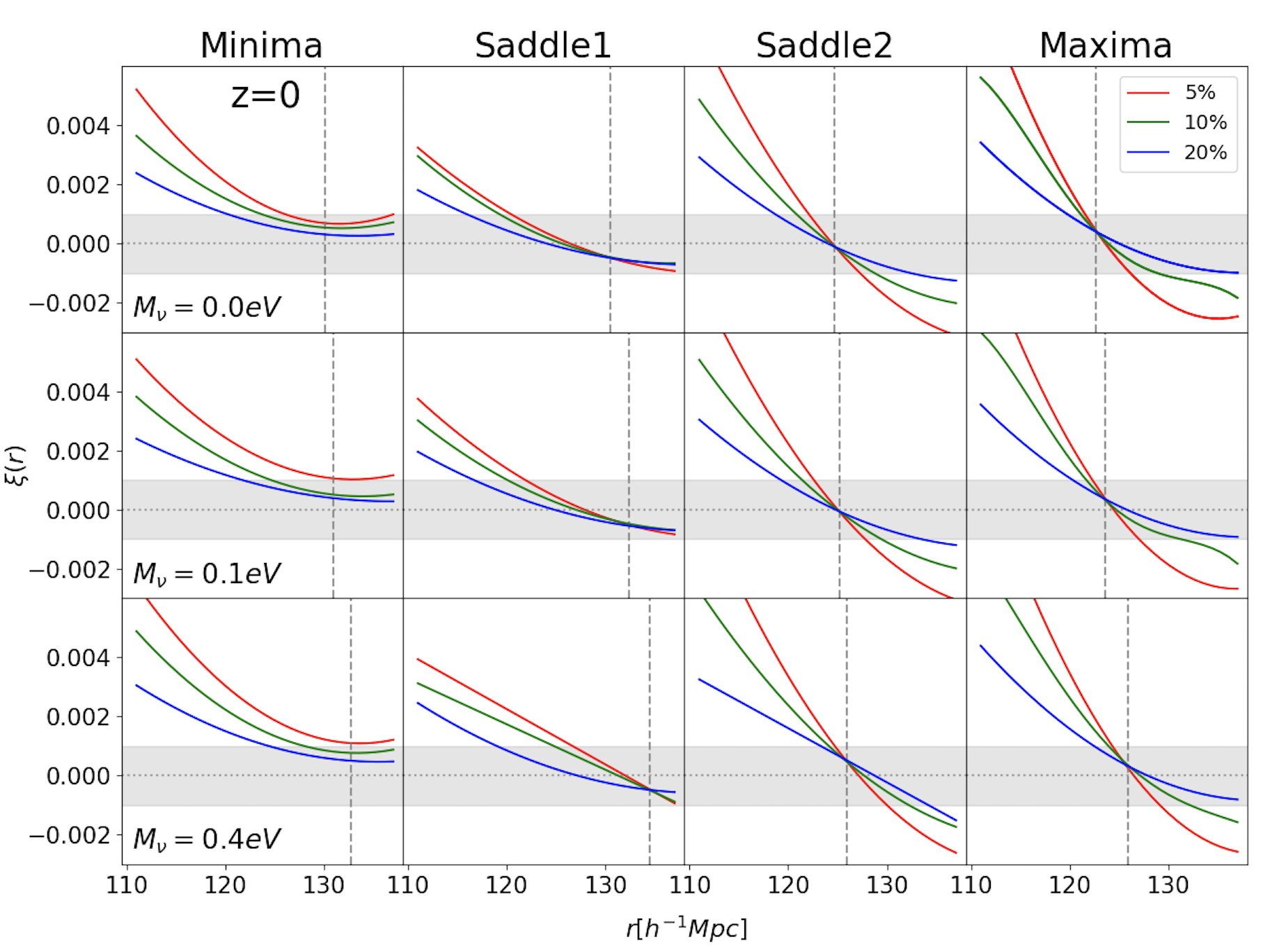}
\caption{Inflection scales of the critical point auto-correlation functions at large separation $r$ presented in Figure \ref{fig_auto_clustering}, indicating a clear sensitivity to massive neutrinos.
In the various panels, measurements of $\xi(r)$ at $z=0$ are performed in the $110h^{-1}{\rm Mpc} < r < 140h^{-1}{\rm Mpc}$ interval, using a refined
binsize of $1h^{-1}{\rm Mpc}$.  
Top panels are for the baseline massless neutrino model, 
middle panels refer to the
scenario with $M_{\nu}=0.1~{\rm eV}$, and
bottom panels show the case of $M_{\nu}=0.4~{\rm eV}$.
Results are averaged over 100 independent 
\texttt{QUIJOTE} realizations per given cosmology. 
The three rarity cuts
are represented by different colors 
as indicated in the panels,
and minima,
wall-type saddles,
filament-type saddles,
and  maxima are reported from left to right.
 Shaded horizontal errorbars indicate
$\pm 0.1\%$ variations in $\xi$.
Remarkably, the
inflection scales (highlighted by the vertical grey  
dashed lines), independently of rarity, are 
altered by a non-zero neutrino mass.}
\label{fig_auto_inflection_details}
\end{figure*}

\begin{figure}
\centering
\includegraphics[angle=0,width=0.49\textwidth]{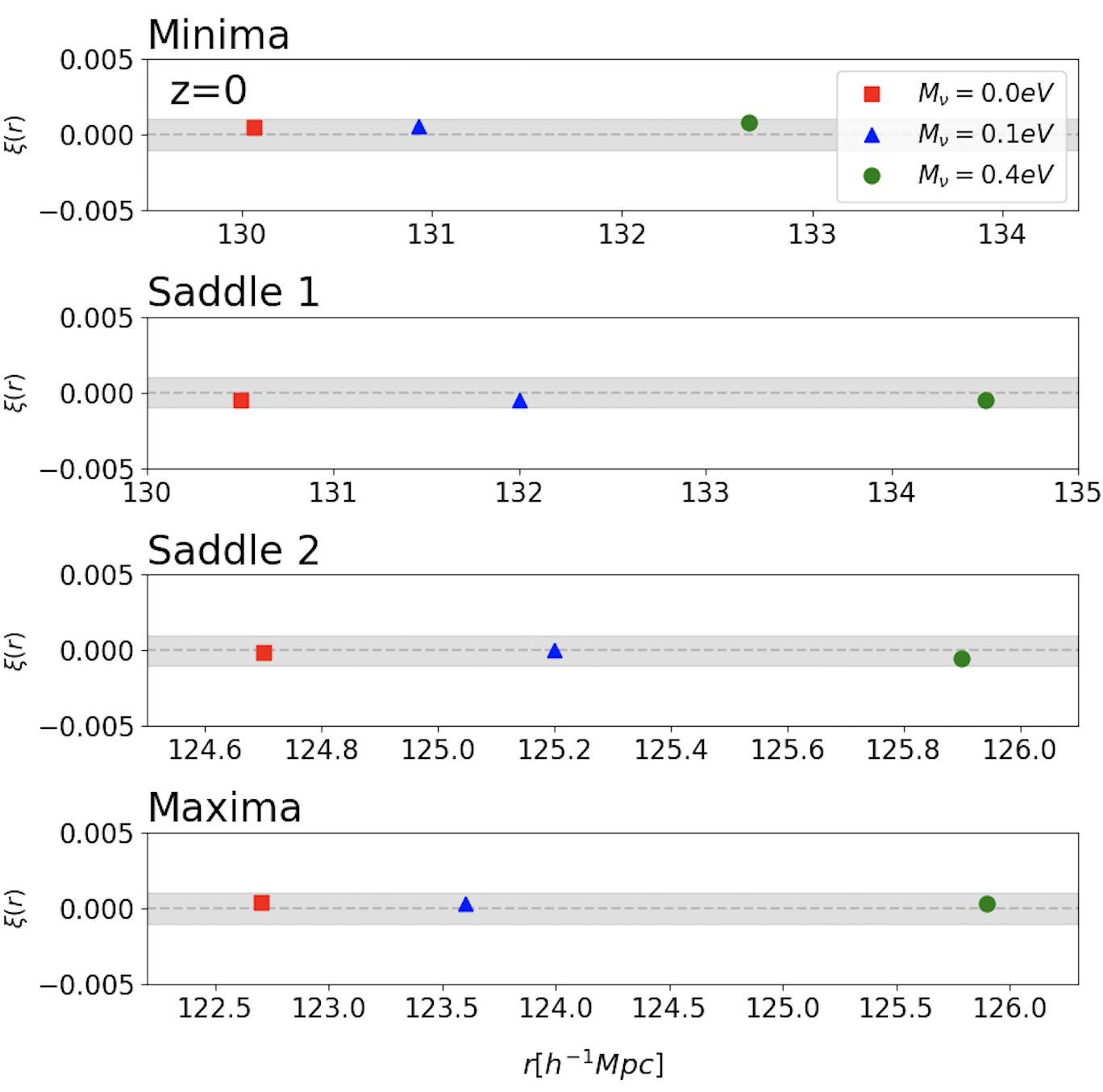}
\caption{Spatial positions of the inflection points of the auto-correlation functions at large scales 
for three different cosmologies, as inferred from Figure \ref{fig_auto_inflection_details}, ordered by 
critical point type (from top to bottom: minima, wall-type saddles, filament-type saddles, maxima). 
Red squares denote $M_{\nu}=0.0~{\rm eV}$, blue triangles are used for $M_{\nu}=0.1~{\rm eV}$,
and green circles show $M_{\nu}=0.4~{\rm eV}$, at $z=0$. Grey horizontal errorbars indicate
$\pm 0.1\%$ variations in $\xi$.  A clear remarkable trend is seen as a function of neutrino mass.}
\label{fig_auto_inflection_scale}
\end{figure}

We then focus on two key features that can be inferred from our two-point correlation measurements in configuration space 
(i.e., Figure \ref{fig_auto_clustering}): namely,
(1) the amplitudes of the various BAO peaks as a function of $\mathcal{R}$, 
and (2) the spatial locations of their corresponding inflection points at large spatial separations. 
The first aspect is quantified via Figures \ref{fig_auto_bao_amplitude} and \ref{fig_auto_bao_amplitude_ratios}, while the second one
is characterized by Figures \ref{fig_auto_inflection_details} and \ref{fig_auto_inflection_scale}.

  
Specifically, Figure \ref{fig_auto_bao_amplitude}
shows  the measured BAO amplitudes ($r^2\xi$)  at $z=0$ along the $x$-axis,
versus their corresponding spatial position 
marked by horizontal grey dashed lines,
for the three cosmological scenarios  
and the three rarity thresholds examined.
From top to bottom, the four panels refer to
minima, 
saddles-1, 
saddles-2, 
and maxima, 
respectively. 
Note in particular that the differences in terms of BAO amplitudes
between a massless neutrino framework and a cosmology with
$M_{\nu}=0.1~{\rm eV}$ are rather small, especially
for the case of minima -- for which their values almost coincide when $\mathcal{R}=10$; 
for this reason, in the top panel the red and blue triangles are slightly displayed along the $y$-direction, 
for the sake of clarity. 
 
Such differences 
are better quantified in Figure \ref{fig_auto_bao_amplitude_ratios}, 
in terms of percentages.
That is, 
we now display relative ratios among BAO amplitudes 
in massive neutrino cosmologies confronted with their correspondent values in a massless neutrino scenario,
as a function of $\mathcal{R}$. Colors refer to the same cosmological 
models considered in Figure  \ref{fig_auto_bao_amplitude},
but now the various symbols are also used to highlight those models,
as reported in the figure.
This visualization is quite useful, as it allows one to readily 
assess the effect of a non-zero neutrino mass on the BAO peaks inferred from the
clustering of critical points in configuration space. In particular, as evident,
departures from analogous measurements carried out in the baseline $M_{\nu}=0.0~{\rm eV}$ framework can reach up to 
$\sim 7\%$ at $z = 0$ 
when $M_{\nu}=0.1~{\rm eV}$,
and  are much more pronounced for higher neutrino mass values. 

From Figures \ref{fig_auto_bao_amplitude} and 
\ref{fig_auto_bao_amplitude_ratios} we 
can deduce a number of significant features. 
First, as in \cite{Shim2021}, we 
find an amplification of the BAO peaks
with rarity: namely, the amplitude of $\xi$ 
is higher (i.e., stronger clustering) 
for lower values of $\mathcal{R}$, regardless of the 
critical point type. 
And BAO features are further amplified by massive neutrinos, as
quantified in the two plots. 
These effects are consistent with -- and generalize --
the findings of  \cite{Peloso2015},
who  characterized the impact of neutrino masses on the shape and height of the BAO peak of the
matter correlation function in real and redshift space, which 
contains relevant cosmological information;
in the nonlinear regime the BAO peak increases with increasing $M_{\nu}$, 
and up to $1.2\%$ at $z=0$ when $M_{\nu} = 0.30~{\rm eV}$, as
reported by the same authors.
Our approach based on critical points
offers a more global \textit{multiscale} perspective, 
since critical points carry remarkable topological 
properties and are good approximations for their corresponding LSS.
Critical points  are also less sensitive to systematic effects, and this is among the reasons
why our auto-correlation BAO peak measurements show that 
departures from the corresponding massless neutrino scenario are more significant (up to $7\%$) 
even when $M_{\nu}=0.1~{\rm eV}$ -- which is a value closer to the current stringent constraints on the summed neutrino mass 
reported in the literature.
Furthermore,  
the spatial position of the BAO
features in the various auto-correlations is robustly defined (i.e.,  $102.5h^{-1}{\rm Mpc}$) 
for all of the abundances considered,  independently of critical point type and neutrino mass 
(hence an excellent standard ruler).
Also, generally
saddle-point statistics are more advantageous
to use for extracting cosmological information 
because their cosmic evolution  is less nonlinear \citep{Gay2012,Shim2021};
for example, the two-point auto-correlations of walls provide
precious information on the characteristic sizes of voids.
Finally, we note that all of the auto-correlation functions go 
to zero at large scales through an inflection point\footnote{We 
use here the same terminology of \cite{Shim2021} to denote  such point, 
namely the spatial location $r_{\rm inf}$ where $\xi(r_{\rm inf}) \equiv 0$ at larger $r$ (i.e., zero-crossing of $\xi$),
although our definition of \textit{inflection scale} is slightly different, as reported in the
main text.}, 
an aspect that we address next. 
 
 
To this end, Figure \ref{fig_auto_inflection_details} 
shows another interesting and remarkable aspect of our analysis:
not only the BAO amplitudes of all of the critical point auto-correlations are
sensitive to massive neutrinos, but also their correspondent \textit{inflection scales}
-- defined
as the spatial positions at large $r$
where the correlation functions $\xi$'s of a given critical point type computed at different 
rarities $\mathcal{R}$ intersect (or are minimally distant), which
also coincide with the spatial locations $r_{\rm inf}$ where $\xi(r_{\rm inf}) \equiv 0$ (i.e., zero-crossings) within errorbars -- 
are altered by a non-zero neutrino mass. 
Specifically,  in Figure \ref{fig_auto_inflection_details}  
we zoom into the $110h^{-1}{\rm Mpc} < r < 140h^{-1}{\rm Mpc}$ interval
and display the various $\xi(r)$'s measurements at $z=0$ 
and $\mathcal{R}=5,10,20$ for
minima, 
wall-type saddles, 
filament-type saddles, 
and  maxima -- from left to right.
All of the correlation function measurements are obtained with a 
`refined zoom-in' technique: in essence, for a given cosmology and within the previously specified spatial interval,
we recompute the two-point auto-correlations shown in Figure \ref{fig_auto_clustering} for the three chosen rarity levels 
using a finer bin size ($1h^{-1}{\rm Mpc}$), and average
the results over 100 independent 
\texttt{QUIJOTE} realizations.  
The inflection points are subsequently determined, 
and in the panels the vertical grey
dashed lines indicate the spatial position of such points,
while the shaded horizontal errorbars highlight variations of $\xi$
by $\pm 0.1\%$.
For minima, the determination of this scale appears more challenging, but it still falls
within the auto-correlation errorbars:
this is likely due to the relatively small resolution of the simulations used in our study,
that impacts underdense regions more severely than other cosmic web components, 
coupled with the fact that minima depart more significantly from linear theory.

The spatial positions of these inflection points at $z=0$ 
are reported in Figure \ref{fig_auto_inflection_scale},
split by corresponding type. 
The shaded horizontal errorbars
indicate the levels where $\xi$ varies by 
$\pm 0.1\%$; as evident, all of the inflection point spatial positions fall
within this range.


\begin{figure*}
\centering
\includegraphics[angle=0,width=0.90\textwidth]{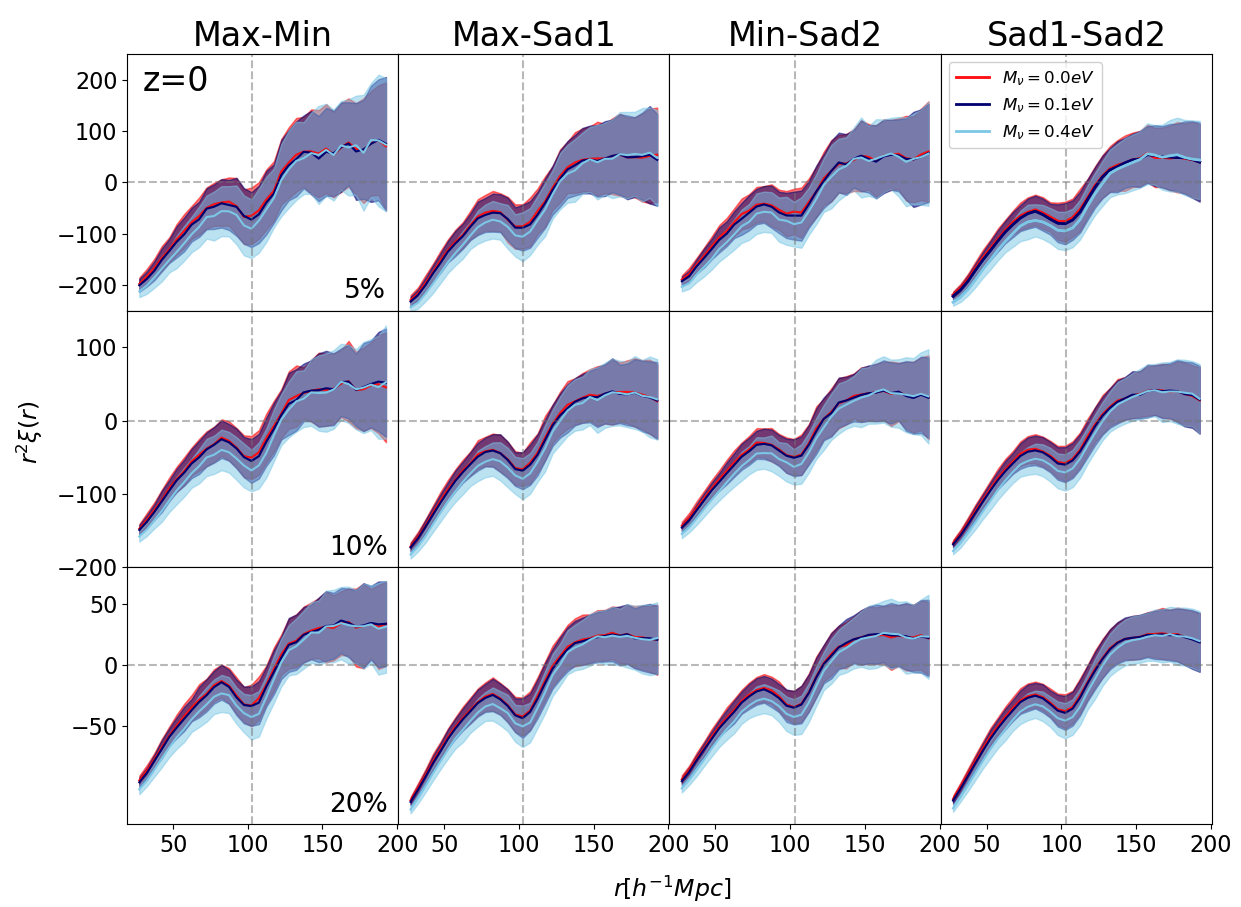} 
\caption{Configuration-space cross-correlations of critical 
points having opposite overdensity sign, as a function of rarity threshold, at $z=0$. 
Colors and line styles same as in  Figure \ref{fig_auto_clustering}.
From left to right, $\mathcal{PV, PW, VF}$, and $\mathcal{WF}$ are displayed when $\mathcal{R}=5$ (top panels),
$\mathcal{R}=10$ (middle panels), and $\mathcal{R}=20$ (bottom panels) -- respectively. 
BAO features now appear as 
dips at $r=102.5h^{-1}{\rm Mpc}$,
with the vertical dashed grey lines marking their exact scale.}
\label{fig_cross_corr_1}
\end{figure*}  


The notable results presented in Figures \ref{fig_auto_inflection_details} 
and \ref{fig_auto_inflection_scale} deserve some further attention.
In particular, 
the interesting fact that such inflection points, 
independently of
rarity, intersect also at the  zero-crossing scale where $\xi(r_{\rm inf}) \equiv 0$ within errorbars  
suggests that their positions can be well-described by standard linear theory, and
we will return on this theoretical aspect in a forthcoming publication.
Furthermore,  these inflection points are also 
quite sensitive to massive neutrinos, with their spatial position increasing with  
augmented neutrino mass -- within the interval [$120-135$]$h^{-1}{\rm Mpc}$. 
As in \cite{Shim2021}, it is then intriguing 
to attempt an analogy, in a multiscale perspective, with the LP of the
correlation function \citep{Anselmi2016}, which is likewise
subject to massive neutrino effects  \citep{Parimbelli2021}.
Figure \ref{fig_auto_inflection_scale} shows indeed
a clear sensitivity to a non-zero neutrino mass: 
through these inflection points it appears feasible to
\textit{simultaneously} 
quantify the perhaps unique signatures of $M_{\nu}$ on the four
key web constituents (i.e., halos, filaments, walls, and 
voids). 


\subsection{Cross-Correlations of Critical Points in Massive Neutrino Cosmologies}    \label{subsec_crosscorr_cp}

\begin{figure}
\centering
\includegraphics[angle=0,width=0.49\textwidth]{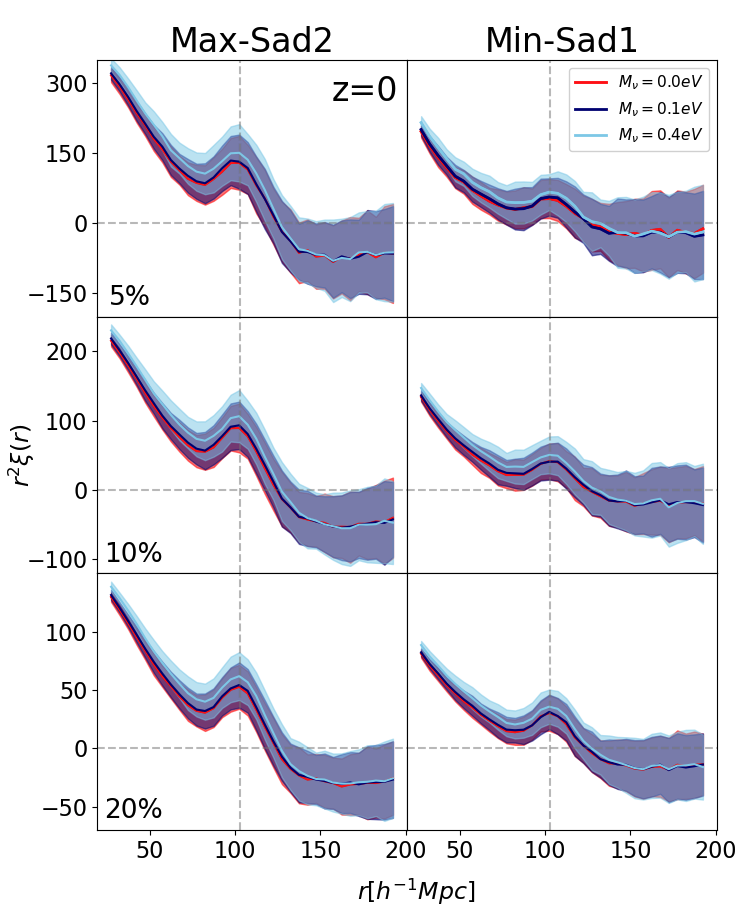}
\caption{Configuration-space cross-correlations of critical 
points characterized by an identical overdensity sign, as a function of rarity threshold, at $z=0$. 
Colors, line styles,  rarity thresholds, and cosmologies 
same as those adopted in Figure \ref{fig_cross_corr_1}. 
Left panels display cross-correlations of overdense critical point pairs ($\mathcal{PF}$), while 
right panels are for underdense critical point cross-correlations ($\mathcal{VW}$). 
Note that the 
shapes of the two-point cross-correlation clustering are
akin to those of auto-correlation measurements (i.e., Figure \ref{fig_auto_clustering}),
since here we consider cross-correlations of similarly biased tracers. 
BAO features appear as  peaks at $r=102.5h^{-1}{\rm Mpc}$,
with the vertical dashed grey lines marking their exact scale.}
\label{fig_cross_corr_2}
\end{figure}

We then move to cross-correlations, and carry out a similar analysis as performed
for the auto-correlation case at $z=0$ in massless and massive neutrino cosmologies, 
for the same rarity thresholds previously considered. 
In general, cross-correlations are helpful in enhancing
 the signal-to-noise ratio (SNR) if used in  combination with auto-correlations,
and in mitigating the impact of systematics.
In what follows, we compute all of 
the possible cross-combinations among extrema and saddle points:
we show their clustering measurements in Figures \ref{fig_cross_corr_1}
and \ref{fig_cross_corr_2}. 
 

Specifically, Figure \ref{fig_cross_corr_1}
displays cross-correlations
among overdense and underdense critical 
points having opposite overdensity sign -- i.e., $\mathcal{PV, PW, VF, WF}$, from left to right, respectively.
The computations are carried out in 
configuration space
at $z=0$, as a function of rarity threshold and for the identical models considered in Figure \ref{fig_auto_clustering}, adopting the  
LS estimator (Equation \ref{eq_cf_ls}). 
Clearly, cross-correlation shapes 
differ from those of auto-correlations (i.e.,  Figure \ref{fig_auto_clustering}). 
In essence, they are 
mirrored with respect to the $x$-axis: 
namely, the clustering is negative (anti-biased, or $\xi(r) <0$ in the $r$-interval of interest) 
until reaching zero at larger spatial separations. 
Hence, at scales relevant for our analysis, these 
cross-correlations are always negative, 
implying that overdense and underdense critical points are anti-correlated.
Therefore,  BAO features are now `reversed' and appear as 
dips (rather than peaks) always at $r=102.5h^{-1}{\rm Mpc}$.
Note that even in this case the spatial positions of the BAO dips are all identical, 
independently of critical point type pair, neutrino mass, and rarity. 
And, similarly to the auto-correlation measurements, 
BAO dips become more pronounced (i.e., showing a steeper negative amplitude)
with increasing neutrino mass. 
As also pointed out by \cite{Shim2021} and \cite{Kraljic2022}, 
anti-clustering arises because
these critical point pairs are oppositely biased tracers of the underlying DM density field.


Figure \ref{fig_cross_corr_2}
shows instead cross-correlations
among overdense \textit{or} underdense critical point pairs (namely, $\mathcal{PF}$ and $\mathcal{VW}$) -- 
characterized by an identical overdensity sign (i.e.,  similarly biased tracers). 
Since now critical point pairs are described by the same
overdensity sign, 
the overall shapes of their two-point clustering are
comparable to those of auto-correlations (see again Figure \ref{fig_auto_clustering}).
Hence, there is no anti-clustering at small separations, BAO peaks (local maxima) are detected at $r=102.5h^{-1}{\rm Mpc}$,
and $\xi(r)$ approaches zero at large spatial separations. 
Also in this case,  BAO peaks are further enhanced by massive neutrinos.
Moreover, we note that $\mathcal{PF}$ cross-correlations
show higher BAO amplitudes than $\mathcal{VW}$ cross-correlations at a fixed rarity threshold. 

   
\begin{figure}
\centering
\includegraphics[angle=0,width=0.48\textwidth]{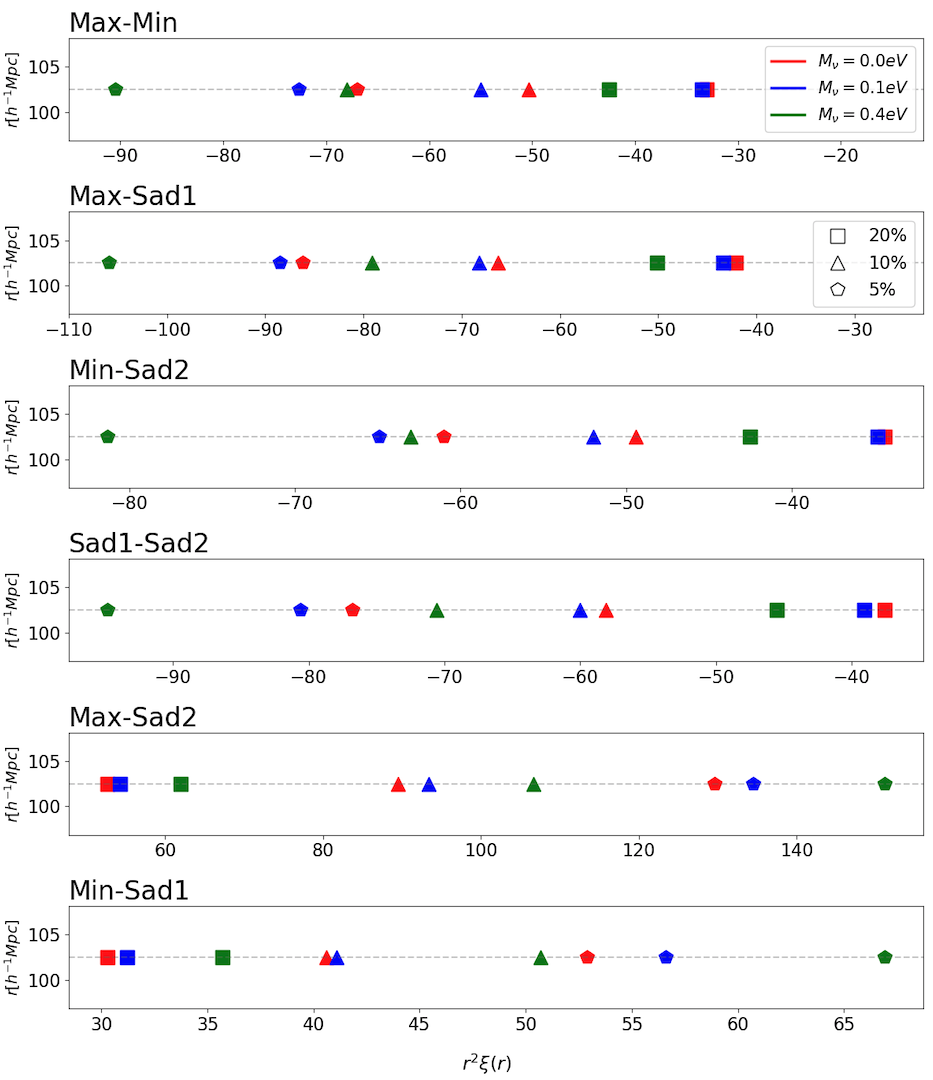}
\caption{BAO dip/peak amplitudes ($r^2\xi$) at $z=0$, versus their correspondent spatial positions ($r=102.5h^{-1}{\rm Mpc}$, 
independent of critical point type pair, neutrino mass, and rarity),
inferred from the cross-correlation measurements presented in Figures \ref{fig_cross_corr_1} and \ref{fig_cross_corr_2}.  
From top to bottom, $\mathcal{PV, PW, VF, WF, PF, VW}$ are shown, respectively. 
Different colors indicate distinct cosmologies, 
while contrasting
symbols represent three choices of rarity thresholds.
A clear trend as a function of neutrino mass can be readily
inferred.}
\label{fig_cross_corr_bao_amplitude}
\end{figure} 


\begin{figure*}
\centering
\includegraphics[angle=0,width=0.90\textwidth]{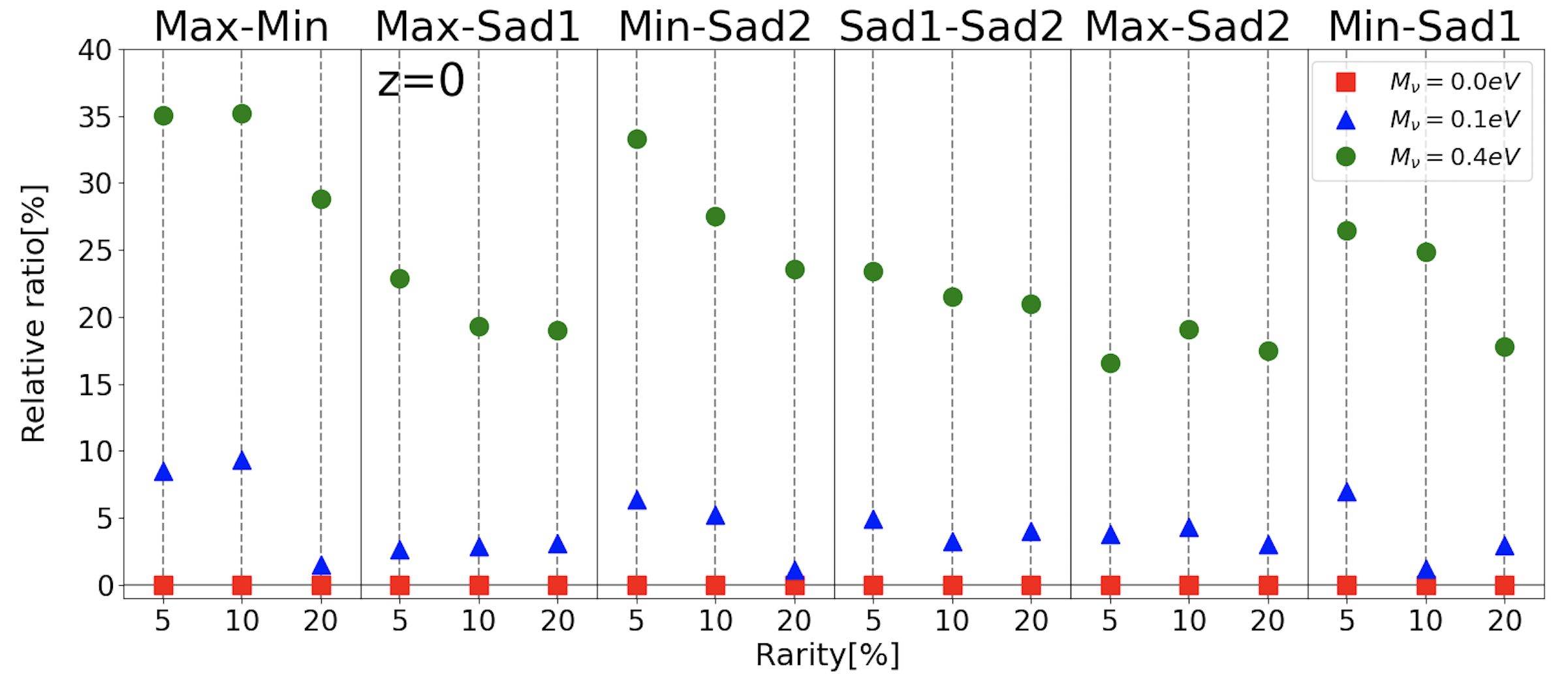}
\caption{BAO dip/peak amplitudes in massive neutrino cosmologies obtained from cross-correlation measurements,  
normalized by their correspondent values in a massless neutrino scenario and expressed in percentages 
as a function of rarity threshold at $z=0$ (with $\mathcal{R}=5,10,20$, respectively). 
Colors and symbols same as in Figure \ref{fig_auto_bao_amplitude_ratios}.
In cross-correlations,  departures from analogous measurements carried out in the baseline $M_{\nu}=0.0~{\rm eV}$ framework 
can be as high as
$\sim 9\%$ at $z = 0$ when $M_{\nu}=0.1~{\rm eV}$.} 
\label{fig_cross_corr_bao_ratios}
\end{figure*} 


Figures \ref{fig_cross_corr_1} and \ref{fig_cross_corr_2}
highlight another interesting aspect of our analysis, 
namely that
the spatial positions of the BAO dips/peaks 
 detected in cross-correlations  are also robustly defined 
 at the scale $r=102.5h^{-1}{\rm Mpc}$  (as for auto-correlations),
independent of critical point type pair, neutrino mass, and rarity.
Moreover, cross-correlations of similarly biased tracers exhibit 
a behavior comparable to those of auto-correlations, 
while BAO features are instead `reversed' and manifest as 
dips for oppositely biased tracers. 
In both scenarios, a non-zero neutrino mass
causes an overall enhancement of the amplitudes of the corresponding peaks/dips
found in cross-correlations, 
much more pronounced for higher neutrino mass values. 
In addition, cross-correlations of minima 
with other critical point types 
show less noisy shapes (i.e., smaller errorbars), if
compared with the auto-correlations of minima alone -- 
which, on the contrary, present the
noisiest correlation shapes. 
Also, $\mathcal{WF}$ cross-correlations
display smooth
and less noisy clustering than extrema cross-correlations. 
Finally, cross-correlations
involving maxima and saddle points 
are also less noisier than
maxima auto-correlations alone, emphasizing 
once more the benefits of cross-correlations. 

\begin{figure*}
\centering
\includegraphics[angle=0,width=0.915\textwidth]{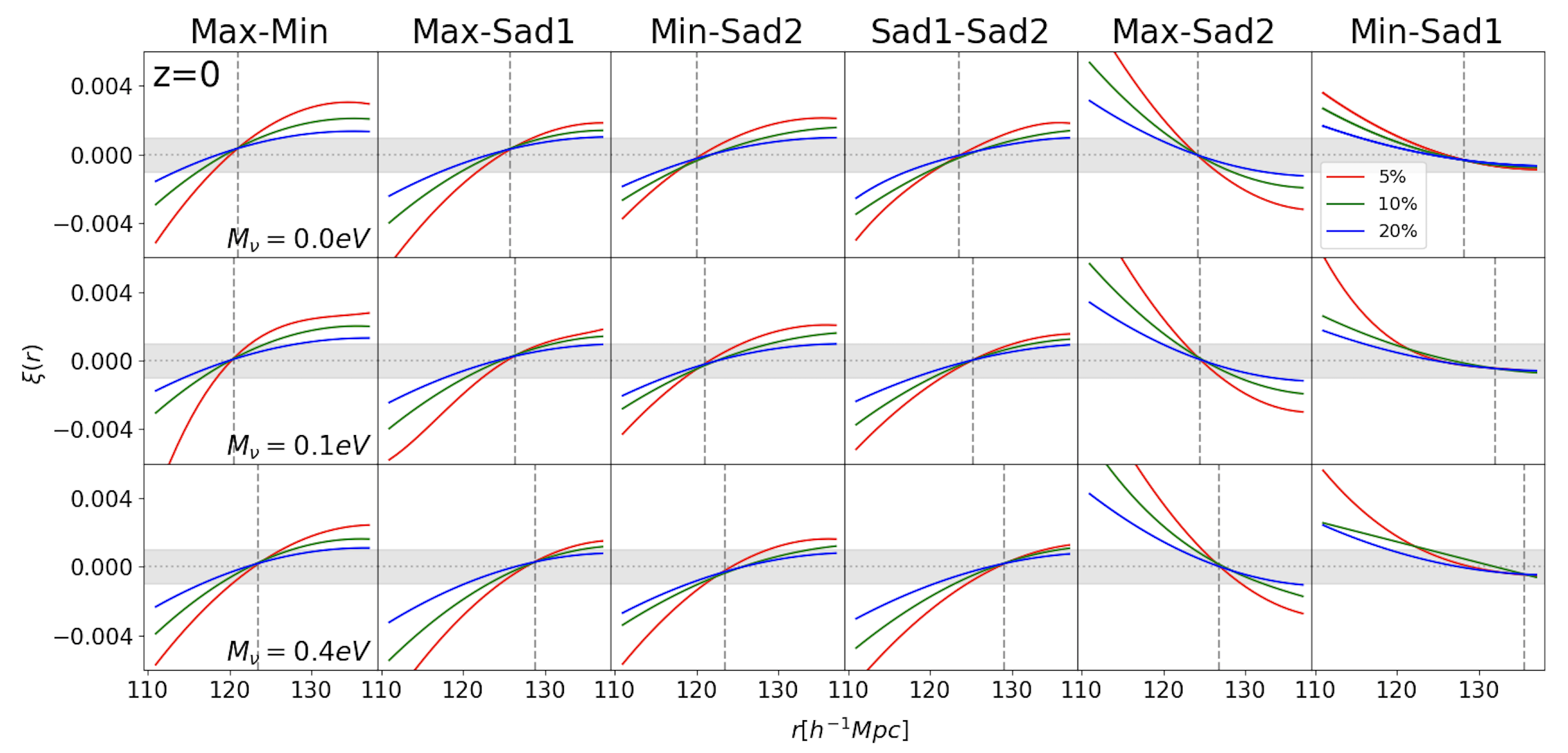}
\caption{Critical point pair cross-correlations: 
determination of the corresponding inflection scales in the interval 
$110h^{-1}{\rm Mpc} < r < 140h^{-1}{\rm Mpc}$, from
averages over 100 independent 
\texttt{QUIJOTE} realizations per cosmology at $z=0$, as specified in the panels. 
From left to right, 
$\mathcal{PV, PW, VF, WF, PF, VW}$ are shown, respectively,
for three rarity thresholds ($\mathcal{R}=5,10,20$)
represented by contrasting colors. 
Shaded horizontal errorbars
display the levels where $\xi$ varies by 
$\pm 0.1\%$. 
In all of the panels, vertical grey
dashed lines indicate the exact positions
of the inflection scales, 
highlighting their  
sensitivity to neutrino mass effects.}
\label{fig_cross_corr_inflection_details}
\end{figure*}
\begin{figure}
\centering
\includegraphics[angle=0,width=0.45\textwidth]{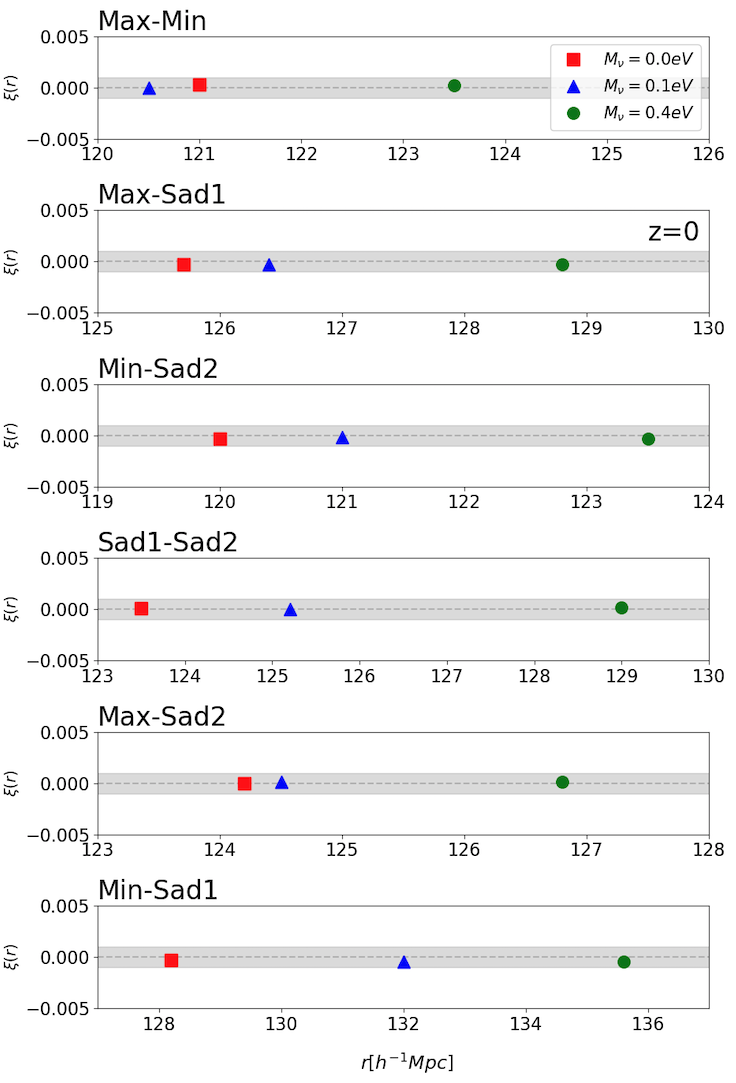}
\caption{Inflection scales of critical point pairs, inferred 
via $\mathcal{PV}$, $\mathcal{PW}$, $\mathcal{VF}$, $\mathcal{WF}$, $\mathcal{PF}$, and $\mathcal{VW}$
cross-correlation measurements 
at $z=0$  (from top to bottom, respectively). Red squares indicate 
$M_{\nu}=0.0~{\rm eV}$,
blue triangles are for $M_{\nu}=0.1~{\rm eV}$,
and green circles denote $M_{\nu}=0.4~{\rm eV}$.  
Shaded horizontal errorbars
highlight
$\pm 0.1\%$ variations in $\xi$: all of the inflection point spatial positions fall
within this range.}
\label{fig_cross_corr_inflection_scale}
\end{figure}

Next, as previously done for auto-correlation measurements, 
we focus on two key aspects
that can be inferred from the two-point cross-correlation estimations in configuration space 
(Figures \ref{fig_cross_corr_1} and \ref{fig_cross_corr_2}): namely,
(1) the amplitudes of the BAO dips/peaks as a function of $\mathcal{R}$, 
and (2) the spatial locations of their corresponding inflection points at large $r$-separations (as defined in the previous section).
The first aspect is addressed in Figures  \ref{fig_cross_corr_bao_amplitude} and \ref{fig_cross_corr_bao_ratios}, 
the second one is quantified via Figures \ref{fig_cross_corr_inflection_details} and \ref{fig_cross_corr_inflection_scale}.


In detail, Figure \ref{fig_cross_corr_bao_amplitude}
shows the 
amplitudes ($r^2\xi$) 
of the BAO dips or peaks  at $z=0$ 
for the various cross-correlations considered 
along the $x$-axis, versus their 
corresponding spatial position 
marked by horizontal grey dashed lines,
for the three cosmologies examined and $\mathcal{R}=5,10,20$.
A clear trend as a function of $M_{\nu}$ 
can be readily
inferred: in essence, BAO dip/peak amplitudes are further enhanced (in absolute value terms)
by the presence of massive neutrinos, and the enhancement is more significant 
with increased neutrino mass. 
Note also that $r^2\xi$ is negative 
in the first four panels since we are considering anti-biased critical points, 
while  $r^2\xi$ is positive for similarly biased tracers. 
 

Figure  \ref{fig_cross_corr_bao_ratios} better quantifies 
such differences
in terms of percentages, in analogy to Figure \ref{fig_auto_bao_amplitude_ratios}
for auto-correlations: namely, we display relative ratios among cross-correlation BAO amplitudes 
in massive neutrino cosmologies, 
normalized by the correspondent values in a massless neutrino scenario,
as a function of $\mathcal{R}$.
As evident from the plot,
departures from analogous cross-correlation measurements carried out in the baseline $M_{\nu}=0.0~{\rm eV}$ framework can reach even 
$\sim 9\%$ at $z = 0$  when $M_{\nu}=0.1~{\rm eV}$,
and  are much more distinct for higher $M_{\nu}$
values.
Interestingly, cross-correlations involving minima
seem to be more effective in distinguishing small neutrino mass
signatures (i.e., when $M_{\nu} =0.1~{\rm eV}$) 
than auto-correlations of minima alone.

 
 Finally, we examine the cross-correlation inflection points at large scales. 
In this respect, 
Figure \ref{fig_cross_corr_inflection_details}
illustrates the remarkable fact that all of
the inflection scales of the
critical point pair cross-correlations
are also sensitive to neutrino mass effects, independently of rarity
(note the analogy with Figure \ref{fig_auto_inflection_details}).
Specifically, 
we consider the $110h^{-1}{\rm Mpc} < r < 140h^{-1}{\rm Mpc}$ interval,
recompute all of the two-point cross-correlations shown in 
Figures \ref{fig_cross_corr_1} and \ref{fig_cross_corr_2}
with a refinement procedure,  
and average the results over 100 independent 
\texttt{QUIJOTE} realizations at $z=0$.  
Inflection points are subsequently determined, as
indicated 
in the panels by the vertical grey
dashed lines, where from left to right we display 
$\mathcal{PV, PW, VF, WF, PF, VW}$, respectively -- 
while the shaded horizontal errorbars
highlight the levels where $\xi$ varies by 
$\pm 0.1\%$. 


The spatial positions of the inflection points 
determined from all of the possible cross-correlations 
at $z=0$ 
are also reported in  Figure \ref{fig_cross_corr_inflection_scale},
split by corresponding pair type.
With the exception of the $\mathcal{PV}$ case when $M_{\nu}=0.1~{\rm eV}$ (top panel),
we detect a clear trend related to the  inflection point positions in neutrino cosmologies:
namely,  the spatial locations of all of the inflection points are amplified by massive neutrinos,
as a function of their mass.  
The apparent disagreement with this trend represented by the 
$\mathcal{PV}$ cross-correlation measurement when $M_{\nu}=0.1~{\rm eV}$ (top panel) 
is likely related to the challenges 
in determining the inflection scale using minima 
(see for comparison the left panels of Figure \ref{fig_auto_inflection_details}),
considering also their selection procedure (i.e., points below rarity threshold, 
see the details in Section \ref{sec_methodology_clustering}).
Remarkably, even in cross-correlations
we find 
an  enhanced sensitivity of the inflection scales to
massive neutrino effects.
 

\subsection{Clustering of Critical Points in Massive Neutrino Cosmologies: Redshift Evolution Effects}  \label{subsec_zevo_cp}

\begin{figure*}
\centering
\includegraphics[angle=0,width=0.75\textwidth]{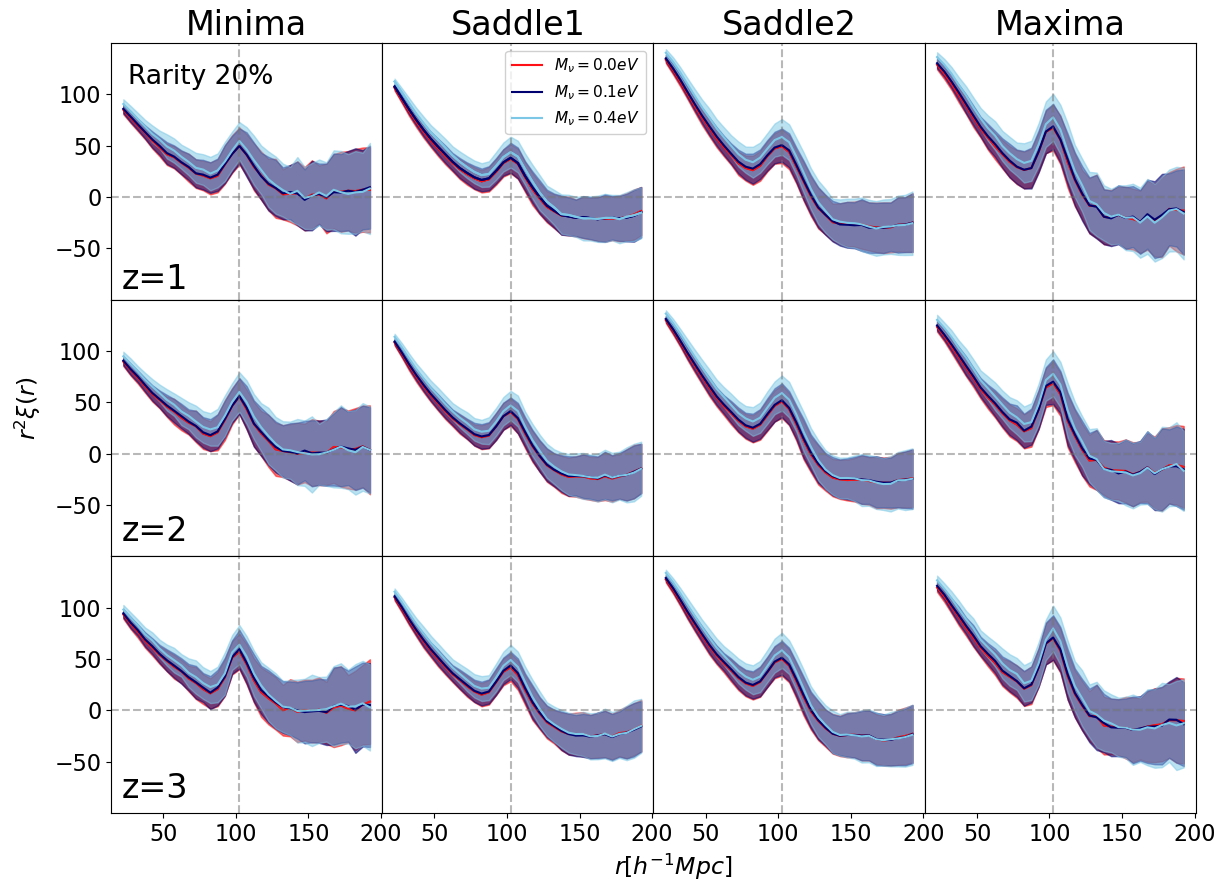}
\caption{Clustering statistics of critical points: redshift evolution at fixed rarity threshold ($\mathcal{R}=20$)
and varying cosmologies (red lines: baseline massless neutrino model; blue lines: $M_{\nu}=0.1~{\rm eV}$;
cyan lines: $M_{\nu}=0.4~{\rm eV}$). Top panels are for $z=1$, middle panels refer to $z=2$, and 
bottom panels are for  $z=3$. Minima,  wall-type saddles, 
filament-type saddles, and 
maxima are represented -- from left to right.} 
\label{fig_z_evolution_example_1}
\end{figure*}
\begin{figure*}
\centering
\includegraphics[angle=0,width=0.75\textwidth]{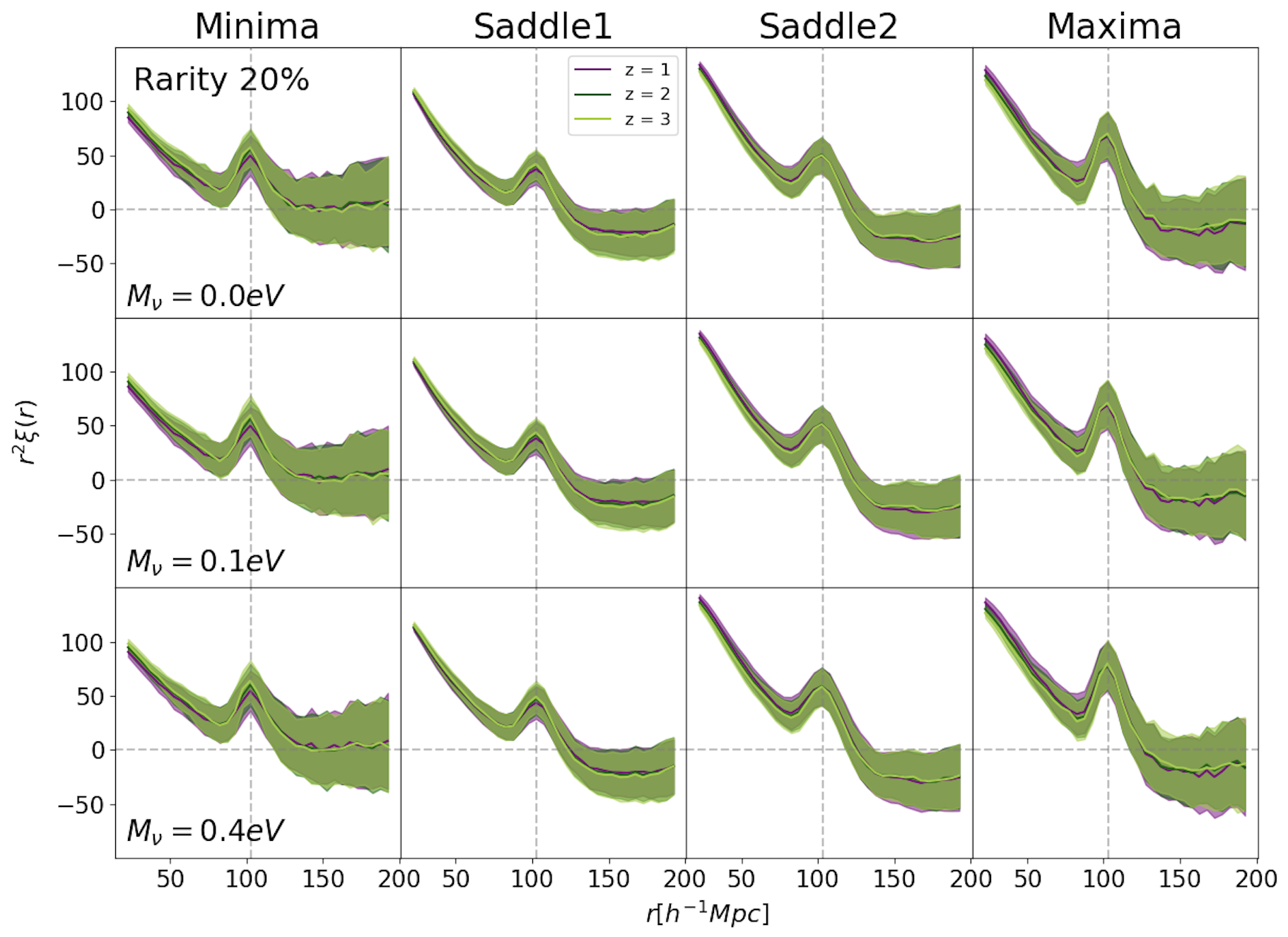}
\caption{Clustering statistics of critical points: redshift evolution at fixed rarity threshold ($\mathcal{R}=20$)
and within a given cosmological framework. Top panels are for the baseline massless neutrino model, 
in the middle panels $M_{\nu}=0.1~{\rm eV}$, and bottom panels display $M_{\nu}=0.4~{\rm eV}$. 
In each panel, contrasting colors represent different redshifts ($z=1,2,3$) -- as specified
by the inset. While at intermediate $r$-separations 
all of the auto-correlations show little evolution with redshift, in general, even if small, 
$z$-evolution is more relevant for the correlations involving the most non-linear critical points (i.e., extrema).}
\label{fig_z_evolution_example_2}
\end{figure*}

Finally, we briefly address redshift evolution effects, in relation to the clustering properties of
critical points. To this end, Figures \ref{fig_z_evolution_example_1} and \ref{fig_z_evolution_example_2} provide some examples for auto-correlations:
the position of the
BAO peaks ($r=102.5h^{-1}{\rm Mpc}$, independent of redshift) is marked by vertical dashed grey lines.  

Specifically, Figure \ref{fig_z_evolution_example_1} displays the 
clustering statistics of critical points in configuration space
at $z=1$ (top panels), $z=2$ (middle panels), and $z=3$ (bottom panels), for
a rarity choice corresponding to $\mathcal{R}=20$ and varying cosmologies.
We adopt the same stylistic conventions as in Figure \ref{fig_auto_clustering}.
In the various panels, we maintain the same $x-y$ scale for a better
visual characterization of the redshift impact in the clustering properties.
At any given redshift and for all of the critical point types,
a non-zero neutrino mass causes an enhancement of the
BAO peak amplitudes: the enhancement is 
more significant for higher neutrino mass values (see for example the $M_{\nu}=0.4~{\rm eV}$ case).
 
To appreciate possible redshift evolution effects more clearly, 
in Figure \ref{fig_z_evolution_example_2} we now show
auto-correlation functions for the same rarity threshold previously examined 
and within a fixed cosmological model -- while we vary the redshift in each panel by considering
$z=1$ (purple), $z=2$ (dark green), and $z=3$ (light green), 
respectively, as specified
by different colors in the figure inset. 
Similarly as in Figure \ref{fig_z_evolution_example_1}, from left to right
we indicate minima,  wall-type saddles, 
filament-type saddles, and 
maxima. 

Around the scale of BAO peaks, 
all of the auto-correlations show small evolution with redshift. 
And, as expected,  
$z$-evolution (although weak) is more relevant for 
the auto-correlations
involving the most non-linear critical points (peaks, voids), while those involving less non-linear critical points (filaments, walls)
appear mostly insensitive to redshift evolution effects. 
In particular, the evolution seems more prominent for minima. 
Moreover, we find a similar trend in cross-correlation measurements. 

Therefore,  as also noted in \cite{Shim2021},
the overall insensitivity to redshift especially for saddle statistics makes the clustering
of critical points a topologically robust alternative to more
standard clustering methods.
 
   
\section{Summary and Outlook}   \label{sec_conclusions}



Determining the neutrino mass scale and type of
hierarchy are primary targets 
for all of the ongoing and upcoming 
large-volume astronomical surveys, as well as
a major goal of future space missions. 
Current limits from cosmology
are already putting considerable pressure on the IH scenario
as a plausible possibility for the neutrino mass ordering 
\citep{Planck2020cosmo,eBOSS2021}, and 
a direct neutrino mass detection, or at least more competitive upper bounds on $M_{\nu}$,
are expected in the next few years from Stage-IV cosmological experiments. 
Moreover, gaining a deeper theoretical understanding
of the impact of massive neutrinos on structure formation particularly at small scales
and on the major observables typically used to characterize  neutrino mass effects
are necessary, to obtain robust
constraints free from systematic biases. 


Traditional methodologies 
for detecting and constraining neutrino mass effects from cosmology
generally rely on a single observable, or on a limited scale-range. 
Examples include CMB gravitational lensing, 
the ISW effect in polarization maps, as well as
a large number of LSS tracers such as
3D galaxy clustering via the 
matter power spectrum, bispectrum,
cosmic shear, galaxy clusters, voids statistics, and much more. 
 

In this work, we pursued a novel and different \textit{multiscale} route, exploiting the
remarkable topological properties of critical points -- 
whose positions and heights at fixed smoothing scale retain
precious cosmological information 
\citep{Sousbie2008,Pogosyan2009,Gay2012,Cadiou2020,Shim2021,Kraljic2022}.
Besides being more robust to systematic effects, and in particular to
non-linear evolution, clustering bias, and redshift space distortions, 
such special points are useful because 
they represent a meaningful and efficient compression
of information of the 3D density field capturing its 
most significant features; thus, they provide
a more fundamental
and global view of the  cosmic web as a whole. 
The overall weak sensitivity to systematics
 ultimately stems from their
close relationship with the underlying topology, as described for example by the genus or Betti numbers.
And, noticeably, 
all of the topological changes of a space occur only at critical points,
and critical points of the density field are
responsible for both the formation \textit{and} destruction of a given feature.


For the first time,  we characterized
their clustering statistics in massive neutrino
cosmologies, addressing the critical point  sensitivity  to small 
neutrino masses
with the goal of identifying (possibly unique) neutrino multiscale signatures
on the corresponding web constituents -- i.e., halos, filaments, walls, and voids. 
We carried out our measurements on
a subset of realizations from the \texttt{QUIJOTE} suite \citep{VillaescusaNavarro2020},
using full snapshots at $z=0,1,2,3$ for 
a choice of representative neutrino masses, as explained in Section \ref{sec_simulations}. 
Exploiting a `\textit{density threshold-based}' methodology,  
described in Section \ref{sec_methodology},
we computed  critical point auto- and cross-correlation functions
in configuration space for a 
series of rarity thresholds, and also characterized redshift evolution effects. 
Our major results, presented in Section \ref{sec_results}
and summarized through Figures \ref{fig_auto_clustering}-\ref{fig_z_evolution_example_2}, 
are focused on BAO scales and are centered
on two key aspects: (1) the multiscale effects of massive neutrinos on the 
BAO peak amplitudes of all of 
the critical point correlation functions above/below rarity threshold, and (2)
the multiscale effects of massive neutrinos on the spatial positions of 
their correspondent correlation function inflection points at large scales.


The main findings and pivotal outcomes of this first work can be summarized as follows: 

\begin{itemize}

\item The BAO scale determined 
from critical point auto- and cross-correlation measurements above/below rarity threshold is 
always robustly detected at  $102.5h^{-1}{\rm Mpc}$. The scale 
coincides with the BAO expected location in the reference cosmology, and it is independent of
critical point type, neutrino mass, and rarity. 
This remarkable aspect is a clear indication that critical points trace the BAO peak similarly to DM, 
halos, and galaxies, and are
faithful representations of their corresponding structures.

\item All of the BAO peaks in auto-correlations are amplified with decreasing rarity: the amplitude of 
$\xi$ is higher for lower values of $\mathcal{R}$,
regardless of the critical point type. Similarly, such
amplifications of the BAO peaks are also found in critical point cross-correlations
characterized by an identical overdensity sign (i.e.,   $\mathcal{PF}$ and $\mathcal{VW}$, similarly biased tracers),
and in the BAO dips of the critical point cross-correlations of 
overdense and underdense regions having opposite overdensity sign (i.e., $\mathcal{PV, PW, VF, WF}$).
Cross-correlations of similarly biased tracers exhibit 
a behavior comparable to those of auto-correlations, 
while BAO features are `reversed' and manifest as 
dips for oppositely biased tracers. Note also that 
the effect of a rarity threshold is also similar to that of smoothing.

\item Massive neutrinos affect the BAO peak amplitudes of all of 
the critical point auto- and cross-correlation
functions above/below rarity threshold. Differences at $z=0$
between a massless neutrino cosmology and a scenario with
$M_{\nu}=0.1~{\rm eV}$  can reach up to
$\sim 7\%$ in auto-correlations, and 
$\sim 9\%$ in cross-correlations. 
BAO peaks/dips become more pronounced (in absolute value terms)
with increasing neutrino mass. 

\item Inflection scales are detected around $125h^{-1}{\rm Mpc}$, both from 
 auto- and cross-correlation measurements. 
Their exact values differ according to the critical point type. 
And, remarkably, inflection scales
are altered by a non-zero neutrino mass -- independently of rarity --
with their spatial position increasing with  
augmented neutrino mass. 

\item Generally, extrema show noisier and less smoother auto- and 
cross-correlation function shapes 
when compared to saddle-point statistics, as 
the time evolution of extrema is more nonlinear than that of saddles. Hence, 
saddle point statistics may be more advantageous 
for extracting cosmological information, and in relation to
neutrino mass constraints.  

\item  Around the BAO scale, 
all of the auto-correlations barely show evolution with redshift. 
Also, as expected,  
$z$-evolution (although weak) is more relevant for 
auto-correlations
involving the most nonlinear critical points (peaks, voids), while those involving less 
nonlinear critical points (filaments, walls)
seem mostly insensitive to redshift evolution effects. 
And,  at any given redshift that we considered and for all of the critical point types,
a non-zero neutrino mass causes an enhancement of the
BAO peak amplitudes, which appears
more significant for higher neutrino mass values. 

\end{itemize}


Our novel approach to neutrino mass effects, based on the statistics of critical points,
offers a \textit{multiscale} perspective, 
since such points carry remarkable topological 
properties and are faithful representation of their corresponding LSS. 
Moreover, critical points  are also less sensitive to systematics, and this is among the reasons
why our auto- and cross-correlation measurements  appear more
sensitive to neutrino signatures than analogous measurements performed on the total
matter correlation function. In this view, 
part of our results can be seen as 
a multiscale generalization of the 
\cite{Peloso2015} findings about the impact of massive neutrinos on BAOs,
and perhaps also as the multiscale extension of the
LP of the spatial correlation function -- 
which is also subject to non-zero neutrino mass effects \citep{ Parimbelli2021}. 
In addition, our technique is complementary to more traditional methods, and can be combined
with such methods to enhance the SNR: we will address this aspect in upcoming publications. 
Hence, our study is particularly relevant for ongoing and future large-volume 
redshift surveys such as DESI and the Rubin Observatory LSST, 
that will provide unique datasets suitable for establishing competitive neutrino mass constraints,
as well as for future high-redshift probes like the 
WHT Extended Aperture Velocity Explorer \citep[WEAVE;][]{Dalton2012}.


To this end, much work still remains to be performed in order
to bring the methodologies presented here to real data applications:
the current study is just the first of a series of investigations that
aim at exploring the sensitivity of critical points and critical events 
to massive neutrinos, and more generally in relation to the dark sector. 
Ongoing efforts are along two major directions: 
towards galaxy clustering at relatively low
redshift, and towards the high-$z$ universe
as mapped for example 
by the 
Ly$\alpha$ forest. 
In forthcoming publications, we will focus on these aspects, and also
characterize the 
relation of our methodology with a persistence-based approach.
Moreover, we will show how
the combined \textit{multiscale} effects 
presented here -- inferred from the clustering of critical points --
can be used  to set 
upper limits on the summed neutrino mass and infer the type of hierarchy. 
 


\begin{acknowledgements}

This work is supported by the National Research Foundation of Korea (NRF) through grant No.
2020R1A2C1005655 funded by the Korean Ministry of Education, Science and Technology (MoEST) 
and by the faculty research fund of Sejong University.
We acknowledge extensive usage of our computing resources (Xeon Silver 4114 master node, 
Xeon Gold 6126 computing nodes, and Lustre file system architecture) at Sejong University. 
G.R. would like to thank Simon White and Volker Springel for kind hospitality 
at the Max-Planck Institute of Astrophysics (MPA Garching) in Summer 2022, where this manuscript was finalized.  
We are also grateful to the referee for insightful feedback and suggestions. 

\end{acknowledgements}

 
 
\bibliography{references}{}
\bibliographystyle{aasjournal} 

 
 
\appendix 

 
 
\section{A. Analysis Technicalities: On Bin Size and Smoothing}  \label{sec_appendix_A}


 \begin{figure*}
\centering
\includegraphics[angle=0,width=0.495\textwidth]{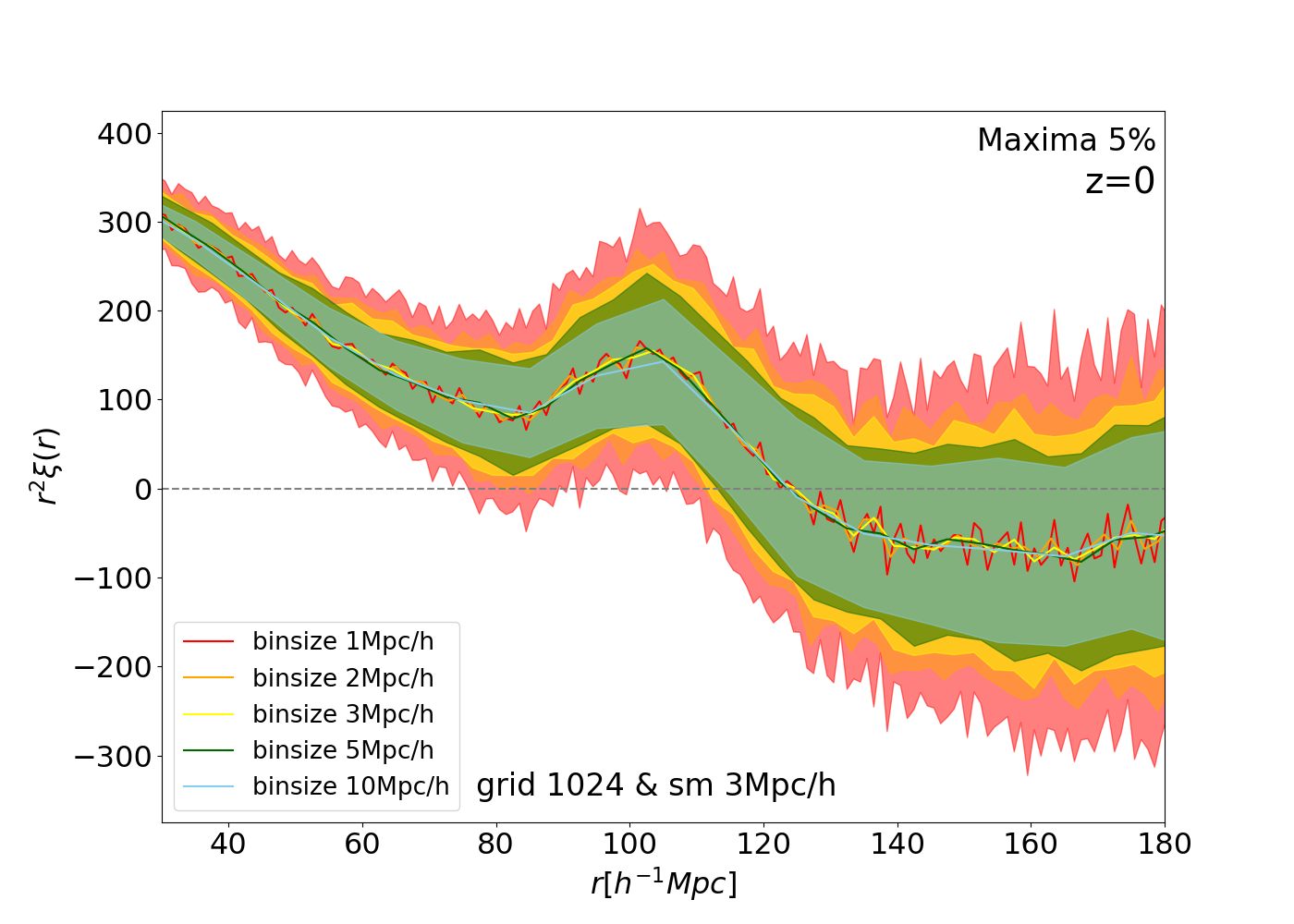}
\includegraphics[angle=0,width=0.495\textwidth]{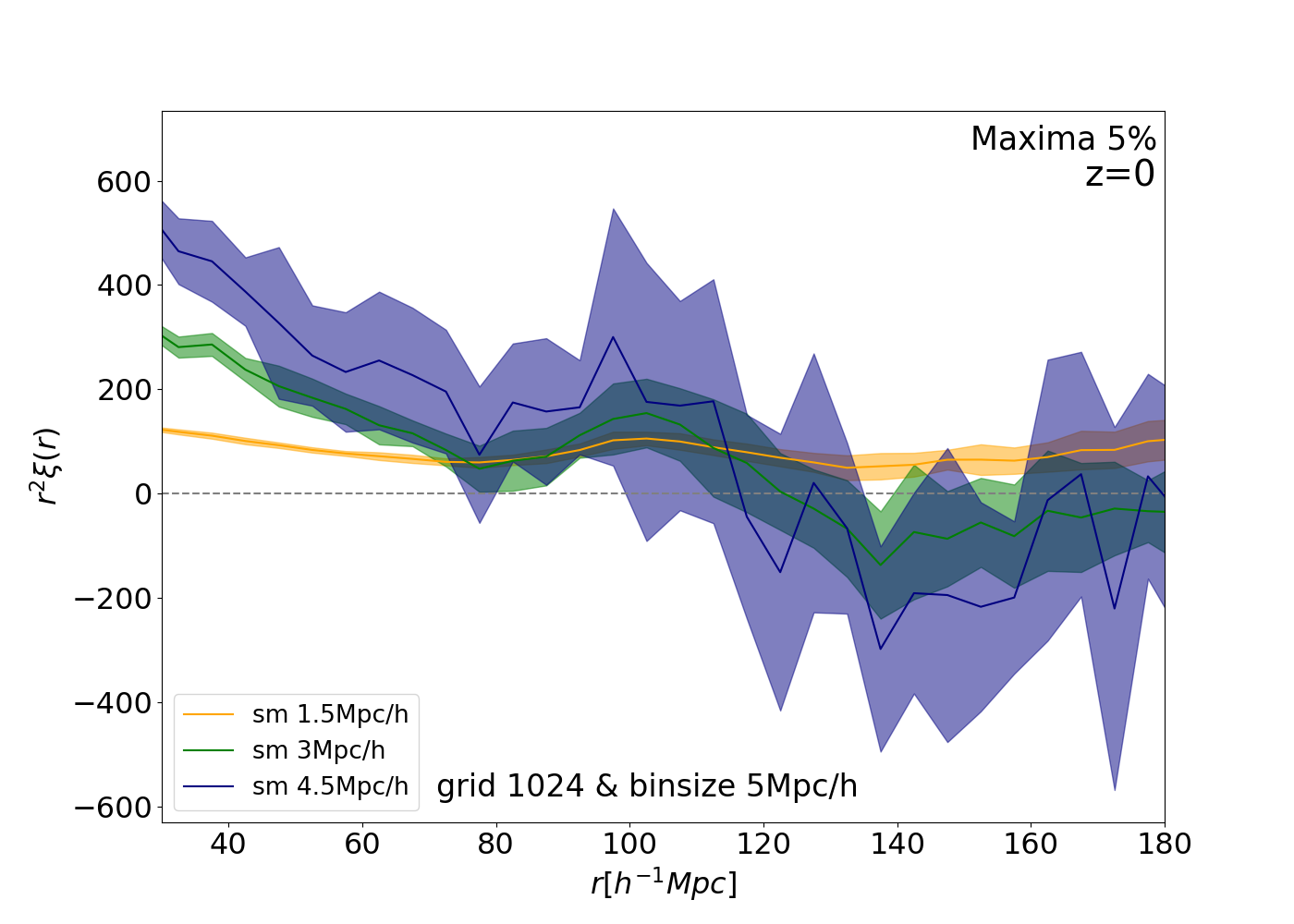}
\caption{Critical point auto-correlation tests related to the bin size choice [left], and to the selection of the 
smoothing scale [right].
In both panels,  we consider only maxima at $z=0$ for a rarity 
$\mathcal{R}=5$ and the number of voxels $N_{\rm grid}=1024$. In particular,  altering the bin size
does not affect the overall shape of the auto-correlation function (including
the location of the BAO peak and the zero-crossing scale), while
smoothing acts similarly to the effect of a rarity choice.}
\label{fig_technicalities}
\end{figure*}  


In this section, we provide more details on the bin size used in the
computations of the auto- and cross-correlation functions presented in the main text, 
and on the smoothing scale adopted.
Specifically, the left panel of Figure \ref{fig_technicalities} shows
the effect of a varying bin size on the critical point auto-correlation function. 
As an example, we consider only maxima at $z=0$ for a rarity threshold 
$\mathcal{R}=5$, when $N_{\rm grid}=1024$ and the smoothing scale is
fixed to $R_{\rm G}=3h^{-1}{\rm Mpc}$. Contrasting colors are used to indicate  
different bin sizes ($r_{\rm bin}$), with $r_{\rm bin} = 1,2,3,5,10$, respectively, in units of $h^{-1}{\rm Mpc}$.  
As evident from the plot, altering the bin size
does not affect the overall shape of the auto-correlation function, including
the location of the BAO peak and the
zero-crossing scale. However, clearly 
smaller bin sizes manifest noisier $\xi$ shapes
(i.e., bigger errorbars). 
Our choice of $5h^{-1}{\rm Mpc}$, represented by the green line with errorbars in the panel, guarantees a
sufficiently smooth $\xi$: this is what 
is typically assumed in surveys like eBOSS for characterizing two-point correlation functions. 
Note also that we adopt a smaller bin size of $1h^{-1}{\rm Mpc}$
in selected regions near inflection points, employing a 
`refinement' technique, in order to accurately determine  their spatial locations
at large scales.
The right panel of Figure \ref{fig_technicalities} 
shows instead the effect of smoothing on correlation function measurements. 
In detail, at a fixed  $N_{\rm grid}=1024$ and for a bin size of
$5h^{-1}{\rm Mpc}$, we compute the two-point auto-correlations
of maxima at $z=0$ when $\mathcal{R}=5$, for different choices of $R_{\rm G}$. 
Contrasting colors represent alternative smoothing scale choices, as specified in the panel,
with $R_{\rm G}=1.5,3.0,4.5$ and $R_{\rm G}$ expressed in $h^{-1}{\rm Mpc}$ units.
As pointed out in the main text, smoothing
acts similarly to the effect of a rarity threshold choice -- see also \cite{Kraljic2022}. 
Namely, an increase in smoothing causes a decrement in 
the number of volume elements along with an increase in bias, and consequently the spatial 
correlation function becomes noisier 
 because of enlarged statistical uncertainties.
In essence, it is the combination of smoothing {\it and} rarity
which is relevant: 
in fact,  smoothing erases small-scale structures, and
too rare events may be too noisy while less rare events will manifest less enhanced characteristic features.  
Generally, decreasing rarity or augmenting smoothing  provides a more significant signal, and
increases the statistical uncertainty. 
Hence, selecting an appropriate smoothing scale along with a suitable rarity is relevant for the
`\textit{density-threshold based}' approach, and the 
choice of rarity is eventually a compromise targeted to the specific problem addressed.
In our specific case, considering three rarities $\mathcal{R}=5,10,20$ and setting $R_{\rm G}=3.0 h^{-1}{\rm Mpc}$
appears to be ideal, since the positions and heights of maxima are
better constrained with smaller smoothing. 

 
 
\section{B. Critical Point Abundance: Useful Tables}  \label{sec_appendix_B}

 
Complementing the information at $z=0$ reported in Tables \ref{table_abundance_cp} and 
\ref{table_abundance_cp_Gauss} that appear in the main text 
(Section \ref{subsec_visualizations_cp}),
we provide here additional details related to redshift evolution effects. Specifically, 
Table \ref{table_abundance_cp_z_evo} contains the 
overall abundance of 
the entire set of critical points at $z=1,2,3$,
classified by type, in the three cosmological frameworks examined in this work (i.e., baseline
massless neutrino model, cosmology with $M_{\nu}=0.1~{\rm eV}$,
and scenario with $M_{\nu}=0.4~{\rm eV}$, respectively).  The associated errorbars are the
corresponding $1\sigma$ variations. 
Table \ref{table_abundance_cp_Gauss} reports instead various 
ratios between the number of peaks ($\mathcal{P}$), 
 voids ($\mathcal{V}$), 
filaments ($\mathcal{F}$), and walls ($\mathcal{W}$) at the same redshifts considered in Table \ref{table_abundance_cp_z_evo},
showing clear 
departures from Gaussian expectations
due to nonlinear evolution -- in addition to the presence of a non-zero neutrino mass.
Note also that
the number of critical points decreases with time (or redshift), although the ratio of extrema over saddles
remains constant.

 
\begin{table*}
\centering
\caption{Total number of critical points at $z=1,2,3$, classified by type, for the three cosmologies considered in this work.}
\doublerulesep2.0pt
 \begin{tabular}{c|c|c|c|c}
\hline \hline   
   & ${\mathbf{M_{\nu}{\rm [eV]}}}$ & ${\mathbf{z=1}}$ & ${\mathbf{z=2}}$ & ${\mathbf{z=3}}$ \\ 
\hline \hline  
&  0.0  &  $232,107 \pm 357$ & $235,768 \pm 373$  &  $237,407 \pm 394$ \\
\cline{2-5}
{\bf Minima ($\mathcal{V}$)} & 0.1    &  $231,667 \pm 368$ & $235,359 \pm 360$ & $237,027 \pm 388$ \\
\cline{2-5}
&  0.4   &  $230,074 \pm 384$ & $233,841 \pm 372$  &  $235,556 \pm 366$ \\
\hline
&  0.0  &  $708,993 \pm 826$ & $724,891 \pm 823$ &  $731,689 \pm 858$ \\
\cline{2-5}
{\bf Saddle 1 ($\mathcal{W}$)} & 0.1  & $707,562 \pm 825$ & $723,567 \pm 797$  & $730,423 \pm 817$ \\
\cline{2-5}
&  0.4   &  $702,347 \pm 833$ & $718,698 \pm 824$ &  $725,859 \pm 827$ \\
\hline
&  0.0   &  $702,078 \pm 846$ & $717,696 \pm 843$ &  $724,529 \pm 860$ \\
\cline{2-5}
{\bf Saddle 2 ($\mathcal{F}$)} & 0.1   & $700,737 \pm 821$ & $716,430 \pm 849$  &   $723,381 \pm 870$  \\
\cline{2-5}
&  0.4   &  $695,547 \pm 819$ & $711,667 \pm 879$ &  $718,871 \pm 879$ \\
\hline
&  0.0   &  $229,913 \pm 348$ & $232,178 \pm 362$ &  $233,099 \pm 389$ \\ 
\cline{2-5}
{\bf Maxima ($\mathcal{P}$)} & 0.1   & $229,546 \pm 350$ &   $231,849 \pm 352$ & $232,776 \pm 397$ \\
\cline{2-5}
&  0.4 &  $228,051 \pm 352$ & $230,434 \pm 350$ &  $231,408 \pm 400$ \\
\hline \hline
\end{tabular}
\label{table_abundance_cp_z_evo} 
\end{table*}

 
\begin{table*}
\centering
\caption{Relevant abundance ratios of critical points at $z=1,2,3$ for the same cosmologies considered in Table \ref{table_abundance_cp_z_evo}, 
highlighting the fact that nonlinear evolution coupled with 
the effects of massive neutrinos (if present) break the
symmetry between overdense and underdense regions.}
\doublerulesep2.0pt
  \begin{tabular}{c|c|c|c|c}
\hline \hline   
   & ${\mathbf{M_{\nu}{\rm [eV]}}}$ &   ${\mathbf{z=1}}$ & ${\mathbf{z=2}}$ & ${\mathbf{z=3}}$ \\ 
\hline \hline  
&  0.0  & $0.9905 \pm 0.0021$& $0.9848 \pm 0.0022$& $0.9819 \pm 0.0023$\\
\cline{2-5}
{\bf  $\mathcal{P/V}$} & 0.1  & $0.9908 \pm 0.0022$ & $0.9851 \pm 0.0021$ & $0.9821 \pm 0.0023$\\
\cline{2-5}
&  0.4   & $0.9912 \pm 0.0023$ & $0.9854 \pm 0.0022$ & $0.9824 \pm 0.0023$ \\
\hline
&  0.0   & $0.9902 \pm 0.0017$ & $0.9901 \pm 0.0016$ & $0.9902 \pm 0.0017$\\
\cline{2-5}
{\bf $\mathcal{F/W}$} & 0.1   & $0.9904 \pm 0.0016$ & $0.9901 \pm 0.0016$ & $0.9904 \pm 0.0016$   \\
\cline{2-5}
&  0.4   & $0.9903 \pm 0.0017$ & $0.9902 \pm 0.0017$& $0.9904 \pm 0.0017$  \\
\hline
&  0.0  & $3.0537 \pm 0.0059$ & $3.0911 \pm 0.0060$ & $3.1082 \pm 0.0064$ \\
\cline{2-5}
{\bf $\mathcal{F/P}$} & 0.1   & $3.0527 \pm 0.0059$ & $3.0901 \pm 0.0060$ & $3.1076 \pm 0.0065$ \\
\cline{2-5}
&  0.4  & $3.0500 \pm 0.0059$ & $3.0884 \pm 0.0060$ & $3.1065 \pm 0.0066$  \\
\hline
&  0.0   & $3.0546 \pm 0.0059$ & $3.0746 \pm 0.0060$ & $3.0820 \pm 0.0063$  \\ 
\cline{2-5}
{\bf $\mathcal{W/V}$} & 0.1  & $3.0542 \pm 0.0060$ & $3.0743 \pm 0.0058$ & $3.0816 \pm 0.0061$  \\
\cline{2-5}
&  0.4   & $3.0527 \pm 0.0063$ & $3.0734 \pm 0.0060$ & $3.0815 \pm 0.0059$ \\
\hline
&  0.0   & $0.9809 \pm 0.0027$ & $0.9750 \pm 0.0027$ & $0.9722 \pm 0.0028$  \\ 
\cline{2-5}
{\bf ($\mathcal{P/W}$)/($\mathcal{V/F}$)} & 0.1   & $0.9813 \pm 0.0027$ & $0.9754 \pm 0.0026$ & $0.9726 \pm 0.0028$  \\
\cline{2-5}
&  0.4  & $0.9816 \pm 0.0028$ & $0.9758 \pm 0.0027$ & $0.9729 \pm 0.0028$ \\
\hline \hline
\end{tabular}
\label{table_abundance_cp_Gauss_z_evo} 
\end{table*}



\end{document}